%% file: main.tex
\documentclass[%
 aip,
 jcp,%
 amsmath,amssymb,floatfix,
 reprint,%
]{revtex4-1}

\usepackage{amsmath,amsthm,latexsym,amssymb,amsfonts,epsfig}

\usepackage[english]{babel}
\usepackage[utf8]{inputenc}			
\usepackage{graphicx, texdraw}   
\usepackage{color}
\usepackage{bm}

\usepackage{lmodern}
\usepackage{mathtools}
\usepackage{mathrsfs} 

\usepackage[margin=0.57in]{geometry}    

\usepackage{enumerate}

\usepackage[colorlinks=true,citecolor=red, linkcolor=blue]{hyperref}

\usepackage{csquotes} 

\usepackage{color}
\usepackage[svgnames,dvipsnames]{xcolor}

\usepackage{amsmath}
\usepackage{amsfonts}
\usepackage{amssymb}
\usepackage{braket}

\usepackage{pdfpages}

\newcommand{\vect}[1]{\boldsymbol{#1}}

\newcommand{\boldNabla}{\boldsymbol{\nabla}}

\newcommand{\boldSigma}{\boldsymbol{\sigma}}

\newcommand{\kS}{\kappa_\mathrm{S}}
\newcommand{\kA}{\kappa_\mathrm{A}}
\newcommand{\kB}{\kappa_\mathrm{B}}
\newcommand{\EB}{E_\mathrm{B}}

\newcommand{\alphaB}{\alpha_\mathrm{B}}

\newcommand{\betaB}{\beta_\mathrm{B}}

\newcommand{\G}{\mathcal{G}}
\newcommand{\R}{\vect{r}}
\newcommand{\Intd}{\mathrm{d }}

\newcommand{\F}{\vect{F}}

\newcommand{\X}{\vect{x}}

\newcommand{\E}{\operatorname{E}}
\newcommand{\bigO}{\mathcal{O}}

 \newcommand{\muP}{\mu^{12}}

\newcommand{\eR}{\vect{e}_r}
\newcommand{\ePhi}{\vect{e}_{\phi}}
\newcommand{\eThe}{\vect{e}_{\theta}}

\newcommand{\gOne}{\vect{g}_1}
\newcommand{\gTwo}{\vect{g}_2}

\newcommand{\Fext}{\vect{F}_2}

\newcommand{\JS}{J_{\mathrm{S}}}

\newcommand{\Gmatr}{\boldsymbol{\mathcal{G}}}

\newcommand{\vStok}{\vect{v}^{\mathrm{S}}}
\newcommand{\vStokcom}{{v}^{\mathrm{S}}}

\newcommand{\xOne}{\vect{x}_1}
\newcommand{\xTwo}{\vect{x}_2}

\newcommand{\vecD}{\vect{d}}
\newcommand{\infSum}{\sum_{n=0}^{\infty}}

\newcommand{\zB}{z_{\mathrm{B}}}

\begin{document}
\title{Hydrodynamic mobility of a solid particle nearby a spherical elastic membrane. I.~Axisymmetric motion}
\author{Abdallah Daddi-Moussa-Ider}
\email{abdallah.daddi-moussa-ider@uni-bayreuth.de}
\author{Stephan Gekle}
\affiliation
{Biofluid Simulation and Modeling, Fachbereich Physik, Universit\"at Bayreuth, Universit\"{a}tsstra{\ss}e 30, Bayreuth 95440, Germany}

\date{\today}

\begin{abstract}
We use the image solution technique to compute the leading order frequency-dependent self-mobility function of a small solid particle moving perpendicular to the surface of a spherical capsule whose membrane possesses shearing and bending rigidities.
Comparing our results with those obtained earlier for an infinitely extended planar elastic membrane, we find that membrane curvature leads to the appearance of a prominent additional peak in the mobility. 
{This peak is attributed to the fact that the shear resistance of the curved membrane involves a contribution from surface-normal displacements which is not the case for planar membranes.}
In the vanishing frequency limit, the particle self-mobility near a no-slip hard sphere is recovered only when the membrane possesses a non-vanishing resistance towards shearing.
We further investigate capsule motion, finding that the pair-mobility function is solely determined by membrane shearing properties.
Our analytical predictions are validated by fully resolved boundary integral simulations where a very good agreement is obtained.


\end{abstract}
\maketitle

\section{Introduction}

Nanoparticles nowadays are widely used in medicine as therapeutic drug delivery agents because of their ability to target specific areas including tumors and inflammation sites \cite{maeda13, chu15}.
Once they are injected into the blood circulation, nanoparticles interact hydrodynamically with neighboring cell membranes in a complex fashion.

In these situations, the Reynolds number is typically very low and a complete description of particle motion is possible via the mobility tensor which gives a linear relation between the particle velocity and the force applied on it.
In the presence of a boundary (interface) the mobility is anisotropic and depends on the distance between the particle and the interface. 
For fluid-solid and fluid-fluid interfaces these mobility tensors have been studied intensively both theoretically \cite{lorentz07, brenner61, goldman67a, goldman67b, lee79, lee80, cichocki98, faucheux94, felderhof05, lauga05, bickel07, swan07, franosch09, blawz10theory, blawz10, lisicki16} and experimentally \cite{dufresne01, CarbajalTinoco_2007, Huang_2007,Choi_2007,  schaffer07, wang09, michailidou09,Kazoe_2011,  lisicki12, rogers12, dettmer14, lisicki14, Wang_2014_diffusion, Misiunas_2015, liu15, traenkle16, benavides16} since quite some time ago.
Due to their relevance as model systems for cell membranes, also elastic interfaces have started to attract some attention recently. 
Here, any motion of the particle causes membrane deformation and a flow is created when the membrane relaxes back to its undeformed state, acting back on the particle motion at a later time.
Accordingly the system possesses a memory and the mobility depends not only on the distance, but also on time or, after temporal Fourier-transformation, on frequency.
Particle motion nearby elastic membranes has been investigated experimentally using optical traps \cite{shlomovitz13, boatwright14, juenger15}, magnetic particle actuation \cite{irmscher12} and quasi-elastic light scattering \cite{mizuno04, kimura05}, where a significant decrease in mobility normal to the cell membrane has been observed similar to that observed near a hard wall.
Particle mobility inside a spherical cell has further been measured by optical microscopy \cite{cervantesMartinez11}.
Setting a particle nearby a cell membrane has been  used in interfacial microrheological experiments as an efficient way to extract membrane's unknown moduli \cite{boatwright14, waigh16}.
Theoretical investigations near elastic interfaces have been carried out using lubrication theory \cite{salez15, saintyves16, rallabandi16}, the point-particle approximation \cite{bickel06, felderhof06, vorobev08, shlomovitz14, bickel14, daddi16, daddi16b, daddi17} and have recently been extended by including higher-order singularities and the hydrodynamic interaction between two particles \cite{daddi16c}. 
All these works considered an infinitely large planar interface which might not always be an appropriate model for a curved cell membrane.
Since their solution technique is based on 2D spatial Fourier transforms \cite{bickel07, Bickel10}, their approach cannot be extended to non-planar interfaces.

In this paper, we therefore employ a different approach based on the image solution technique to compute the frequency dependent mobility of a small particle  moving perpendicular to an initially spherical elastic object (which can be a cell, a capsule or a vesicle) whose membrane exhibits resistance towards shearing and bending.
The method has originally been introduced by Fuentes and coworkers \cite{fuentes88, fuentes89} who investigated the hydrodynamic interactions between two unequal viscous drops when the interparticle gap is of the order of the diameter of the smaller one.

The remainder of the paper is organized as follows.
In Sec.~\ref{sec:singularitySolution}, we compute the flow field by expressing the solution of the fluid motion as a multipole expansion.
In Sec.~\ref{sec:particleMobility}, we give analytical expressions of the particle frequency-dependent self-mobility in terms of infinite series, nearby idealized membranes with shearing-only or bending-only rigidities.
The motion of the capsule is studied in Sec.~\ref{sec:capsuleMotion}, finding that the pair-mobility function depends only on membrane shearing properties.
A comparison between theoretical predictions and numerical simulations is provided in Sec.~\ref{sec:comparisonWithBIM} where a very good agreement is obtained.
A conclusion summarizing our results is offered in Sec.~\ref{sec:conclusions}.
The technical details are relegated to the appendices.

\section{Singularity solution}\label{sec:singularitySolution}

\begin{figure}  
	\centering
	\def\svgwidth{0.95\linewidth}
	\large
	\includegraphics{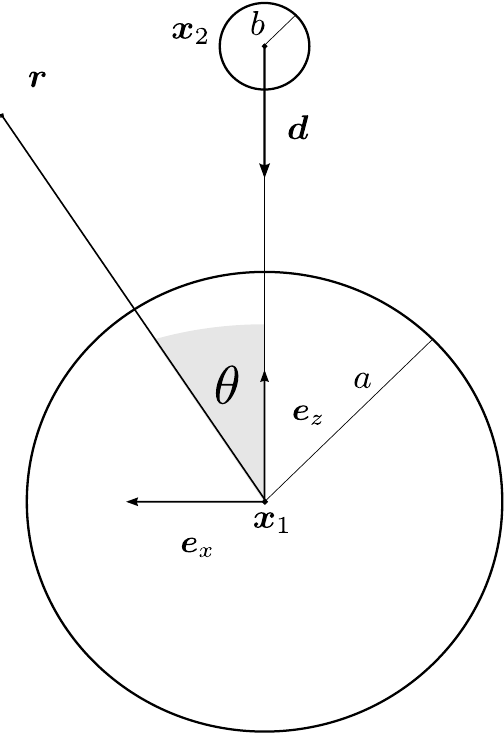}
	\caption{Illustration of the system setup.  
	A small solid spherical particle of radius $b$ positioned at $\xTwo = R \vect{e}_z$ nearby a large spherical capsule of radius $a$.
	In an axisymmetric configuration, the force is directed along the unit vector $\vecD \equiv -\vect{e}_z$.
	}
	\label{illustration}
\end{figure}

In this section, we derive the image solution for a point-force acting nearby a spherical capsule of radius $a$.
We will use the term \enquote{capsule} to denote a general soft object including cells or vesicles.
The origin of spherical coordinates is located at $\xOne$, the center of the capsule.
An arbitrary time-dependent point-force $\vect{F}$ is acting at $\xTwo = R \vect{e}_z$
(see Fig.~\ref{illustration} for an illustration of the system setup.)
The problem is thus equivalent to solving the forced Stokes equations 
\begin{align}
 \eta \boldNabla^2 \vect{v} - \boldNabla p + \vect{F} \delta (\X - \xTwo) &= 0 \, , \label{Stokes:Momentum} \\
 \boldNabla \cdot \vect{v} &= 0 \, , \label{Stokes:Continuity}
\end{align}
for the fluid outside the capsule and 
\begin{align}
 \eta \boldNabla^2 \vect{v}^{(i)} - \boldNabla p^{(i)} &= 0 \, , \label{Stokes:Momentum_Inside} \\
 \boldNabla \cdot \vect{v}^{(i)} &= 0 \, , \label{Stokes:Continuity_Inside}
\end{align}
inside.
Here $\vect{v}$ and $p$ denote the flow velocity and the pressure outside the capsule, and the superscript~$(i)$ denote the corresponding interior fields.
For simplicity, the fluid is assumed to have the same dynamic viscosity $\eta$ everywhere.

We therefore need to solve Eqs.~\eqref{Stokes:Momentum} through \eqref{Stokes:Continuity_Inside} for the boundary conditions imposed at the membrane equilibrium position $r=a$,
 \begin{align}
  [v_\theta] &= 0 \, , \label{BC:v_phi} \\
  [v_r] &= 0 \, , \label{BC:v_r} \\
  [\sigma_{\theta r}] &= \Delta f_\theta^{\mathrm{S}} + \Delta f_\theta^{\mathrm{B}} \, , \label{BC:sigma_r_phi} \\
  [\sigma_{rr}] &= \Delta f_r^{\mathrm{S}} + \Delta f_r^{\mathrm{B}} \, , \label{BC:sigma_r_r}
 \end{align}
where the notation $[w] := w(r = a^{+}) - w(r = a^{-})$ represents the jump of a given quantity $w$ across the membrane.
Here we assume axisymmetry such that all azimuthal components vanish.
Throughout the remainder of this paper, all the lengths will be scaled by the capsule radius $a$ unless otherwise stated.
For convenience, the transition rules to physical quantities are summarized in appendix~\ref{appendix:transformationEquations}.
The non-vanishing components of the fluid stress tensor are expressed in spherical coordinates as \cite{kim13}
\begin{subequations}
 \begin{align}
   \sigma_{\theta r} &= \eta \left( v_{\theta,r} - \frac{v_\theta}{r} + \frac{v_{r,\theta}}{r} \right) \, , \label{sigma_r_phi} \\
   \sigma_{rr} &= -p + 2\eta v_{r,r} \, , \label{sigma_r_r}
 \end{align}
\end{subequations}
where comma in indices denotes a spatial partial derivative.
Note that Eqs.~\eqref{BC:v_phi} and \eqref{BC:v_r} represent the natural continuity of the flow field across the membrane, whereas Eqs.~\eqref{BC:sigma_r_phi} and \eqref{BC:sigma_r_r} are the discontinuity of the normal-tangential and normal-normal components of the fluid stress tensor at the membrane.
Here $\Delta f_\theta$ and $\Delta f_r$ are the meridional and radial traction where the superscripts S and B stand for the shearing and bending related parts, respectively.
As derived in Appendix \ref{appendix:membraneMechanics}, according to the Skalak model \cite{skalak73} the linearized traction due to shearing elasticity reads
\begin{subequations}
 \begin{align}
  \Delta f_{\theta}^{\mathrm{S}} &= -\frac{2\kS}{3} \bigg( (1+2C) u_{r,\theta} + (1+C) u_{\theta,\theta\theta} \notag \\
		  &+  (1+C)u_{\theta,\theta} \cot\theta -  \left( (1+C) \cot^2 \theta + C \right) u_\theta \bigg)   \, ,   \\
  \Delta f_r^{\mathrm{S}} &=  \frac{2\kS}{3} (1+2C) \left( 2u_r + u_{\theta,\theta} + u_{\theta} \cot \theta \right)   \, .  \label{radialTractionJumpShearing}
 \end{align}
 \label{tractionJumpShear}
\end{subequations}

The traction jump due to bending resistance can be derived from the Helfrich model \cite{helfrich73} or by assuming a linear constitutive relation for the bending moments \cite{pozrikidis01jfm}.
For small deformations, both formulations are equivalent \cite{bendingReview} leading to the traction (cf.~appendix~\ref{appendix:membraneMechanics})
\begin{subequations}
 \begin{align}
  \Delta f_{\theta}^{\mathrm{B}} &=  {\kB} \left( \left(1-\cot^2\theta\right)u_{r,\theta} + u_{r,\theta\theta} \cot\theta + u_{r,\theta\theta\theta} \right)   \, ,  \\
  \Delta f_{r}^{\mathrm{B}} &= {\kB} \bigg( \left( 3\cot\theta+\cot^3\theta \right) u_{r,\theta} -  u_{r,\theta\theta}\cot^2\theta \notag \\
  &+  2 u_{r,\theta\theta\theta}\cot\theta + u_{r,\theta\theta\theta\theta} \bigg)  \, . 
 \end{align}
 \label{tractionJumpBend}
\end{subequations}
Here $\vect{u} (\theta) = u_r (\theta) \eR + u_\theta (\theta) \eThe$ denotes the membrane displacement vector, related to the fluid velocity by the no-slip relation at $r=1$ by
\begin{equation}
\left.  \vect{v}\right|_{r=1} = \frac{\Intd \vect{u}}{\Intd t}   \, , \label{no-slip}
\end{equation}
which can thus be written in temporal Fourier space as $\vect{v} = i\omega \, \vect{u}$ evaluated at $r=1$.   
The membrane parameters $\kS$ and $\kB$ are the shearing and bending moduli, respectively, and $C$ is the Skalak parameter defined as the ratio between area expansion modulus $\kA$ and shear modulus $\kS$.
An unscaled version of the above equations in physical units can be obtained by applying the rules given in appendix~\ref{appendix:transformationEquations}.

Our resolution approach is based on the image solution method proposed by Fuentes \textit{et al.} \cite{fuentes88} who computed the axisymmetric motion of two viscous drops in Stokes flow.
Accordingly, the exterior fluid velocity can be written as a sum of two contributions,
\begin{equation}
 {v}_i = \vStokcom_i + {v}_i^{*} \, ,
\end{equation}
where $\vStokcom_i := \G_{ij} (\vect{x} - \xTwo) {F}_j $ is the velocity field induced by a point-force acting at $\xTwo$ (cf.~equation~\eqref{stokeslet_at_X2}) in an infinite medium, i.e. in the absence of the capsule and ${v}_i^{*}$ is the image system required to satisfy the boundary conditions at the capsule membrane.

Now we briefly sketch the main resolution steps. 
First, the velocity $\vStok$ due to the Stokeslet acting at $\xTwo$ is written in terms of spherical harmonics which are transformed afterward into harmonics based at~$\xOne$ via the Legendre expansion.
Second, the image system solution $\vect{v}^{*}$ is expressed as multipole series at $\xOne$ which subsequently is rewritten in terms of spherical harmonics centered at~$\xOne$.
Third, the solution inside the capsule $\vect{v}^{(i)}$ is expressed using Lamb's solution \cite{lamb32} also written in terms of spherical harmonics at~$\xOne$.
The last step consists of determining the series expansion coefficients by satisfying the boundary conditions at the membrane surface stated by Eqs.~\eqref{BC:v_phi} through~\eqref{BC:sigma_r_r}.

\subsection{Stokeslet representation}

We begin with writing the Stokeslet acting at $\xTwo$,
\begin{equation}
 \vStokcom_i =  \G_{ij} F_j = \frac{1}{8\pi\eta} \left( F_i \frac{1}{s} + F_j (\vect{x} - \xTwo)_i {\nabla_2}_j \frac{1}{s} \right) \, , 
 \label{stokeslet_at_X2}
\end{equation}
where $s:= |\vect{x}-\xTwo|$.
Here ${\nabla_2}_j := {\partial }/{\partial  {x_2}_j}$ denotes the nabla operator taken with respect to the singularity position $\xTwo$.
Using Legendre expansion, the harmonics based at $\xTwo$ can be expanded as
\begin{equation}
 \frac{1}{s} = \infSum \frac{r^{2n+1}}{R^{n+1}} \frac{(\vecD \cdot \boldNabla)^n}{n!} \frac{1}{r} \, , 
\end{equation}
where the unit vector $\vecD := ( \xOne - \xTwo)/R = -\vect{e}_z$, $\R = \X - \xOne$ and $r := |\R|$.
Moreover, we denote by $\varphi_n$ the harmonic of degree~$n$, related to the Legendre polynomials of degree~$n$ by \cite{abramowitz72}
\begin{equation}
 \varphi_n (r, \theta) := \frac{(\vecD \cdot \boldNabla)^n}{n!} \frac{1}{r} = \frac{1}{r^{n+1}} P_n (\cos \theta) \, .
\end{equation}

For the axisymmetric case, the force is exerted along the unit vector $\vecD$ and can be written as $\vect{F} = F \vecD$.
By making use of the identities
\begin{equation}
 \boldNabla_2 \frac{1}{R^{n+1}} = \frac{n+1}{R^{n+2}} \, \vecD \, , \quad  (\vecD \cdot \boldNabla_2)\, \vecD = 0 \, ,  \label{twoIdentities}
\end{equation}
Eq.~\eqref{stokeslet_at_X2} can therefore be written as
\begin{equation}
 \begin{split}
  \vStok &= \frac{F}{8\pi\eta} \bigg[  \infSum (n+2)\frac{r^{2n+1}}{R^{n+1}} \, \vecD \, \varphi_n  \\
         &+ \infSum (n+1) \frac{r^{2n+1}}{R^{n+2}} \, \R  \varphi_n \bigg] \, . \label{vStok_withDependentHarmonics}
 \end{split}	
\end{equation}

Hence, the Stokeslet is written in terms of harmonics based at $\xOne$.
Note that the terms with $\vecD\,\varphi_n$ in Eq.~\eqref{vStok_withDependentHarmonics} are not independent harmonics.
For their elimination, we shall use the following recurrence property \cite{fuentes88}
\begin{equation}
 \begin{split}
  \vecD\,\varphi_n &= \frac{1}{2n+1} \bigg( \boldNabla \varphi_{n-1} - r^2 \boldNabla \varphi_{n+1} \\
                   &-(2n+3) \R \, \varphi_{n+1} \bigg) \, , 
 \end{split}
 \label{eliminateDependency}
\end{equation}
leading after substitution into Eq.~\eqref{vStok_withDependentHarmonics} to
\begin{equation}
 \begin{split}
 \vStok &= \frac{F}{8\pi\eta}  \sum_{n=1}^{\infty} \Bigg[ \left( \frac{n+3}{2n+3} \frac{r^{2n+3}}{R^{n+2}} - \frac{n+1}{2n-1} \frac{r^{2n+1}}{R^n} \right) \boldNabla \varphi_n \\
                      &+ \left( (n+1) \frac{{r^{2n+1}}}{R^{n+2}} - \frac{(n+1)(2n+1)}{2n-1} \frac{r^{2n-1}}{R^n} \right) \R  \varphi_n \Bigg] \, .
 \end{split}
 \label{stokeslet_Finalize()}
\end{equation}
Note that the terms with $n=0$ cancel so that the summation starts from $n=1$.

\subsection{Image system representation}

Next, we write the image system solution following a multipole expansion approach as 
\begin{equation}
 \begin{split}
  {v}_i^{*} &= \frac{F \, d_j}{8\pi\eta}  \infSum \bigg[ A_n \frac{(\vecD \cdot \boldNabla)^n}{n!} \G_{ij}(\R) \\
            &+ B_n \frac{(\vecD \cdot \boldNabla)^n}{n!} \boldNabla^2 \G_{ij} (\R)  \bigg] \, , \label{imageSystemRepresentation}
 \end{split}
\end{equation}
where the solution form is assumed as a result of the system axisymmetry \cite{fuentes88} with the constants $A_n$ and $B_n$ to be determined by the boundary conditions.
By making use of the identity
\begin{equation}
 \frac{(\vecD \cdot \boldNabla)^n}{n!} \G_{ij} (\R) = \delta_{ij} \varphi_n - r_i \frac{\partial \varphi_n}{\partial x_j} - d_i \frac{\partial \varphi_{n-1}}{\partial x_j} \, .  \notag
\end{equation}
together with
\begin{equation}
 \boldNabla^2 \G_{ij} (\R) = -\frac{\partial^2}{\partial x_i \partial x_j} \frac{2}{r} \, , \notag
\end{equation}
the image solution can be written as 
\begin{equation}
 \begin{split}
  \vect{v}^{*} &= -\frac{F}{8\pi\eta} \infSum \Bigg[ A_n \Big( (n-1)\vecD \, \varphi_n + (n+1) \R  \varphi_{n+1}\Big) \notag \\
              &+ 2(n+1)B_n \boldNabla \varphi_{n+1}  \Bigg] \, .
 \end{split}
\end{equation}

Further, the elimination of the dependent harmonics $\vecD\, \varphi_n$ is readily achieved using Eq.~\eqref{eliminateDependency}.
Shifting the index to start the sum from $n=1$, we finally obtain
\begin{equation}
 \begin{split}
  \vect{v}^{*} &=  \frac{F}{8\pi\eta} \sum_{n=1}^{\infty} \Bigg[ 
		   \bigg( \frac{n-2}{2n-1} r^2 A_{n-1} - \frac{n}{2n+3}A_{n+1} \\ 
		   &- 2nB_{n-1} \bigg) \boldNabla \varphi_n 
		   -\frac{2(n+1)}{2n-1} A_{n-1} \R \varphi_n  
                   \Bigg] \, .
 \end{split}
               \label{imageSolution_Finalize()}
\end{equation}

\subsection{Solution inside the capsule}

For the flow field inside the capsule, we use Lamb's general solution \cite{cox69, misbah06}, which can be expressed in terms of interior harmonics based at $\xOne$ as \cite{fuentes88}
\begin{equation}
 \begin{split}
  \vect{v}^{(i)} &= \frac{F}{8\pi\eta} \sum_{n=1}^{\infty} \Bigg[ 
                  a_n \bigg( \frac{n+3}{2}r^{2n+3} \boldNabla \varphi_n  \\
                  &+ \frac{(n+1)(2n+3)}{2} r^{2n+1} \R \varphi_n \bigg) \\
                  &+ b_n \left( r^{2n+1} \boldNabla \varphi_n + (2n+1)r^{2n-1} \R \varphi_n \right)
		  \Bigg] \, .
 \end{split}
 \label{LambSolution_Finalize()}
\end{equation}

The determination of the series coefficients outside the capsule $A_n$ and $B_n$ and inside the capsule $a_n$ and $b_n$ is achieved by applying the boundary conditions at the capsule membrane.
This will be subject of the next subsections.

\subsection{Determination of the series coefficients}

Hereafter, for the sake of  completeness, we shall state explicitly the expressions of the projected velocity components onto the radial and tangential directions.
For this aim, we make use of the following identities for the projection onto the radial direction,
\begin{subequations}
 \begin{align}
  \eR \cdot \boldNabla \varphi_n &= -\frac{n+1}{r} \varphi_n \, , \\
  \eR \cdot \R \varphi_n &= r \varphi_n \, .
 \end{align}
\end{subequations}
For the projection onto the tangential direction, we make use of
\begin{equation}
  \eThe \cdot \R \varphi_n = 0 \, .
\end{equation}
We further define
\begin{equation}
 \psi_n := \eThe \cdot \boldNabla \varphi_n   = \frac{1}{r} \frac{\partial \varphi_n}{\partial \theta} \, . \label{psi_n}
\end{equation}

From Eq.~\eqref{stokeslet_Finalize()}, the radial and tangential components of the Stokeslet solution follow forthwith.
We obtain
 \begin{align}
  v_r^{\mathrm{S}} &= \frac{F}{8\pi\eta} \sum_{n=1}^{\infty} 
                                    \left[ \frac{n(n+1)}{2n+3}\frac{r^{2n+2}}{R^{n+2}} - \frac{n(n+1)}{2n-1} \frac{r^{2n}}{R^{n}} \right] \varphi_n \, , \label{v_r_S} \\
  v_{\theta}^{\mathrm{S}} &= \frac{F}{8\pi\eta}  \sum_{n=1}^{\infty} \left[ \frac{n+3}{2n+3} \frac{r^{2n+3}}{R^{n+2}} - \frac{n+1}{2n-1} \frac{r^{2n+1}}{R^n} \right] \psi_n \, . \label{v_phi_S}
 \end{align}

Similar, from Eq.~\eqref{imageSolution_Finalize()} we obtain the components of the image solutions as
 \begin{align}
  v_r^{*} &= \frac{F}{8\pi\eta} \sum_{n=1}^{\infty} \bigg[ -\frac{n(n+1)}{2n-1} r A_{n-1} + \frac{n(n+1)}{2n+3}\frac{A_{n+1}}{r} \notag \\
          &+ 2n(n+1) \frac{B_{n-1}}{r} \bigg] \varphi_n \, , \label{v_r_Im} \\
  v_{\theta}^{*} &= \frac{F}{8\pi\eta} \sum_{n=1}^{\infty} \left[ \frac{n-2}{2n-1} r^2 A_{n-1} - \frac{n A_{n+1}}{2n+3}  - 2nB_{n-1} \right] \psi_n \, . \label{v_phi_Im}
 \end{align}

From Eq.~\eqref{LambSolution_Finalize()}, the components of the flow field inside the capsule read
 \begin{align}
  v_r^{(i)} &= \frac{F}{8\pi\eta} \sum_{n=1}^{\infty} \left[ \frac{n(n+1)}{2} r^{2n+2} a_n + n r^{2n} b_n \right] \varphi_n \, , \label{v_r_Inside} \\
  v_\theta^{(i)} &=  \frac{F}{8\pi\eta} \sum_{n=1}^{\infty} \left[ \frac{n+3}{2} r^{2n+3} a_n + r^{2n+1} b_n \right] \psi_n \, . \label{v_phi_Inside}
 \end{align}

\subsubsection*{Pressure field}

In order to proceed later, we need to express the pressure field in terms of a multipole expansion.
The form of the pressure $p$ in the exterior fluid follows from the general solution of the axisymmetric Laplace equation in spherical coordinates as
\begin{equation}
 p = \frac{F}{8\pi} \sum_{n=1}^{\infty} (S_n+ Q_n r^{2n+1}) \varphi_n \, . \notag
\end{equation}
Since the form of the velocity field is known from Eqs.~\eqref{v_r_S}-\eqref{v_phi_Im}, the coefficients $S_n$ and $Q_n$ can be related to the coefficients of the velocity field by using Eq.~\eqref{Stokes:Momentum} leading to 
\begin{equation}
 S_n = -2nA_{n-1} \, , \quad Q_n = \frac{2(n+1)}{R^{n+2}} \, . \label{pressureCoeffs}
\end{equation}

Inside the capsule, all harmonics of negative order which lead to a singularity at $r=0$ need to be discarded reducing the form of the pressure to
\begin{equation}
 p^{(i)} = \frac{F}{8\pi} \sum_{n=1}^{\infty} p_n r^{2n+1} \varphi_n \, . \notag
\end{equation}
Using Eqs.~\eqref{Stokes:Momentum_Inside}, \eqref{v_r_Inside} and \eqref{v_phi_Inside} we find
\begin{equation}
 p_n = {(n+1)(2n+3)} a_n \, .
\end{equation}

\subsubsection{Continuity of velocity}

After substituting Eqs.~\eqref{v_r_S} through \eqref{v_phi_Inside} into Eqs.~\eqref{BC:v_phi} and \eqref{BC:v_r}, the continuity of the tangential and radial fluid velocity components across the membrane leads to the two following equations  
\begin{subequations}
 \begin{align}
  \frac{n(n+1)}{2} \, a_n + n b_n &= -\frac{n(n+1)}{2n-1} \, A_{n-1}+ \frac{n(n+1)}{2n+3} \, A_{n+1} \notag \\
                               &+ 2n(n+1) B_{n-1} + \frac{n(n+1)}{2n+3} \frac{1}{R^{n+2}} \notag \\ 
                               &- \frac{n(n+1)}{2n-1} \frac{1}{R^n} \, , \notag \\
  \frac{n+3}{2} \, a_n + b_n &= \frac{n-2}{2n-1} \, A_{n-1} - \frac{n \, A_{n+1}}{2n+3}  - 2n B_{n-1} \notag \\
                               & + \frac{n+3}{2n+3} \frac{1}{R^{n+2}} - \frac{n+1}{2n-1} \frac{1}{R^n} \, , \notag
 \end{align}
\end{subequations}
which can be solved for the coefficients $a_n$ and $b_n$ to obtain
 \begin{align}
  a_n &=  A_{n-1}-\frac{2n+1}{2n+3} \, A_{n+1} -2(2n+1) B_{n-1} \notag \\
      &+ \frac{2}{2n+3} \frac{1}{R^{n+2}} \, , \label{a_n} \\
  b_n &=  - \frac{(n+1)(2n+1)}{2(2n-1)} \, A_{n-1} + \frac{n+1}{2} \, A_{n+1} \notag \\
      & + (n+1)(2n+3) B_{n-1} - \frac{n+1}{2n-1} \frac{1}{R^n} \, . \label{b_n}
 \end{align}

\subsubsection{Discontinuity of the stress tensor}

Expressions for $A_n$ and $B_n$ can be determined from the discontinuity of the traction across the membrane.
In order to assess the effect of shearing and bending on particle self-mobility, we shall consider in the following shearing and bending effects separately.

\paragraph{Shearing contribution}

Here we consider an idealized membrane with a shearing-only resistance, such as a typical artificial capsule \cite{rachik06}.
After setting $\Delta f_r^{\mathrm{B}} = \Delta f_\theta^{\mathrm{B}} = 0$ in the traction jump equations given by Eqs.~\eqref{BC:sigma_r_phi} and \eqref{BC:sigma_r_r}, we readily obtain
\begin{subequations}
   \begin{align}
    \left[ v_{\theta,r}\right] &= 
     -\alpha \bigg( (1+2C) v_{r,\theta} + (1+C) \left( v_{\theta,\theta\theta} + v_{\theta,\theta} \cot\theta \right)\notag \\
            &- \left.   \left( (1+C) \cot^2 \theta + C \right) v_\theta \bigg) \right|_{r=1}  \, , \label{jump_v_Phi_Shear} \\
    \bigg[ \frac{p}{\eta} \bigg]  &= 
    {\left. \alpha (1+2C) v_{r,r} \right|_{r=1}} \, , \label{jump_v_r_Shear}
   \end{align}
\end{subequations}
where $i \alpha := 2\kS/(3\eta \omega)$ {upon using the incompressibility equation}
\begin{equation}
 \frac{2v_r}{r}+v_{r,r}+\frac{v_{\theta,\theta} + v_\theta \cot \theta}{r} = 0 \, .  \notag
\end{equation}
It follows immediately that $[v_{r,r}] = 0$. 
Furthermore, note that $[v_{r,\theta}]=0$.

Continuing, we proceed first by substituting the expressions of the velocity components given by Eqs.~\eqref{v_r_S}-\eqref{v_phi_Inside} into the tangential traction jump Eq.~\eqref{jump_v_Phi_Shear} and replacing $a_n$ and $b_n$ with their expressions given by Eqs.~\eqref{a_n} and \eqref{b_n}, respectively.
For the determination of the unknown coefficients, we multiply both equation members by $\psi_m \sin\theta$ and integrate over the polar angle $\theta$ between 0 and $\pi$.
By making use of the following orthogonality properties
\begin{equation}
 \int_{0}^{\pi} \psi_{m}  \psi_{n}  \sin\theta \Intd \theta = \frac{2n(n+1)}{2n+1} \frac{\delta_{mn}}{r^{2n+4}} \, , \label{orthoProp_psi_1}
\end{equation}
and
\begin{equation}
  \begin{split}
    & \int_{0}^{\pi} \psi_{m} \left(\psi_{n,\theta\theta} + \psi_{n,\theta} \cot \theta - \psi_{n} \cot^2 \theta \right) \sin\theta \Intd \theta \\
               &=  -\frac{2n(n+1)(n^2+n-1)}{2n+1} \frac{\delta_{mn}}{r^{2n+4}} \, . \label{orthoProp_psi_2}
  \end{split}
\end{equation} 
the resulting equation reads 
\begin{widetext}
 \begin{equation}
  \begin{split}
  (2n+1) \Big( 2(2n+3) B_{n-1} - A_{n-1} + A_{n+1} \Big)  = \\
                                              -\alpha \bigg( (1+2C)n (n+1)
  \left( \frac{A_{n+1}}{2n+3} -\frac{A_{n-1}}{2n-1}  + 2 B_{n-1} 
                               -\frac{1}{2n-1} \frac{1}{R^{n}} + \frac{1}{2n+3} \frac{1}{R^{n+2}} \right)   \\
 + \left( \frac{n-2}{2n-1}A_{n-1} - \frac{n}{2n+3} A_{n+1} - 2n B_{n-1} - \frac{n+1}{2n-1} \frac{1}{R^n} + \frac{n+3}{2n+3} \frac{1}{R^{n+2}}  \right) 
  \left( 1 - (1+C)n(n+1) \right) \bigg) \, ,                                  
  \end{split}
  \label{jump_f_Phi_Finalize()}
\end{equation}
\end{widetext}
for $n\ge 1$.
Next, we write a similar equation for the normal traction jump Eq.~\eqref{jump_v_r_Shear}.
After substituting the velocity and the pressure into Eq.~\eqref{jump_v_r_Shear},
multiplying both members by $\varphi_m \sin\theta$ and employing the orthogonality properties
\begin{equation}
 \int_{0}^{\pi} \varphi_n \varphi_m \sin\theta \Intd \theta = \frac{2}{2n+1} \frac{\delta_{mn}}{r^{2n+2}} \, . \label{orthoProp_phi_1}
\end{equation}
and
\begin{equation}
 \begin{split}
  \int_{0}^{\pi} \varphi_m \left( \varphi_{n,\theta\theta} + \varphi_{n,\theta} \cot\theta \right) \sin\theta \Intd \theta \\
  = -\frac{2n(n+1)}{2n+1} \frac{\delta_{mn}}{r^{2n+2}} \, . \label{orthoProp_phi_2}
 \end{split}
\end{equation}
we get after replacing $a_n$ and $b_n$ with their corresponding expressions 
\begin{widetext}
 \begin{equation}
 \begin{split}
  -2(2n+3)(2n+1)(n+1)B_{n-1} + (2n^2+7n+3) A_{n-1} - (2n^2+3n+1)A_{n+1} \\
  = \alpha (1+2C) n (n+1)
  \bigg( 
  -\frac{n}{2n-1} A_{n-1} + \frac{n+2}{2n+3} A_{n+1} + 2(n+2) B_{n-1} 
  +\frac{n-1}{2n-1} \frac{1}{R^{n}} - \frac{n+1}{2n+3} \frac{1}{R^{n+2}} 
  \bigg) \, ,
 \end{split}
 \label{jump_f_R_Finalize()}
\end{equation}
\end{widetext}
for $n\ge 1$.

The equations \eqref{jump_f_Phi_Finalize()} and \eqref{jump_f_R_Finalize()} form a closed linear system of equations, amenable to immediate resolution using the standard substitution method.
From Eq.~\eqref{jump_f_Phi_Finalize()}, $B_{n-1}$ can be expressed in terms of $A_{n-1}$ and $A_{n+1}$.
We obtain
\begin{equation}
 \begin{split}
  B_n &= -\frac{A_{n+2}}{4n+10} + \frac{1}{2G} \bigg( \frac{G' A_n}{2n+1} + \frac{\alpha G_3}{2n+5} \frac{1}{R^{n+3}} \\
      &- \frac{\alpha G_1}{2n+1} \frac{1}{R^{n+1}} \bigg) \, , 
 \end{split}
 \label{B_n_Shearing}
\end{equation}
for $n \ge 0$, 
where we defined 
\begin{subequations}
 \begin{align}
  G   &:= (C+1)\alpha n^3 + [(6C+5)\alpha + 4] n^2 \notag \\
      &+ [(11C+7)\alpha+16]n + (6C+3)\alpha +15 \, , \notag \\
  G'  &:= \alpha (1+C)n^3+[(4C+3)\alpha+4]n^2 \notag \\
      &+ [(5C+1)\alpha+8]n+(1+2C)\alpha+3 \, , \notag \\
  G_1 &:= (C+1) n^3 + (3C+4) n^2 + 2(C+2) n \, , \notag \\
  G_3 &:= (1+C) n^3 + (5C+6) n^2+ (8C+10) n \notag \\
      &+ (4C+2) \, . \notag
 \end{align}
\end{subequations}

Next, by substituting the expression of $B_{n-1}$ into Eq.~\eqref{jump_f_R_Finalize()}, we obtain the general term for $A_{n}$ as
\begin{equation}
   A_{n} = \frac{\alpha n (n+2)}{K}
 \bigg( \frac{K_3 }{R^{n+3}}  - \frac{K_1 }{R^{n+1}} \bigg) \, ,
  \label{A_n_Shearing}
\end{equation}
for $n \ge 0$ where 
\begin{subequations}
 \begin{align}
  K   &:= 8(C+1)\alpha n^5 + [(4C+2)\alpha^2+60(C+1)\alpha+32]n^4 \notag \\
      &+[(24C+12)\alpha^2+172(C+1)\alpha +192]n^3 \notag \\
      &+[(44C+22)\alpha^2+234(C+1)\alpha+400] n^2 \notag \\
      &+[(24C+12)\alpha^2+(150C+138)\alpha+336]n \notag \\
      &+ (36C+18)\alpha + 90 \, , \notag \\
  K_1 &:= 4(C+1)n^4+[(4C+2)\alpha+20C+28]n^3 \notag \\
      &+ [(22C+11)\alpha+31C+75]n^2 \notag \\
      &+ [(36C+18)\alpha+15C+93]n + (18C+9)\alpha + 45 \, , \notag \\
  K_3 &:= 4(C+1)n^4+[(4C+2)\alpha+20C+28]n^3 \notag \\
      &+ [(18C+9)\alpha+35C+71]n^2 \notag \\
      &+[(20C+10)\alpha+25C+71]n + (6C+3)\alpha +6C +21 \, . \notag
 \end{align}
\end{subequations}

The general term for $B_n$ can then be obtained by substituting the expressions of $A_{n}$ and $A_{n+2}$ determined from Eq.~\eqref{A_n_Shearing} into Eq.~\eqref{B_n_Shearing}.

In particular, for $\alpha \to \infty$ (achieved either by taking an infinite membrane elastic modulus or by considering a vanishing frequency) we recover the hard-sphere limit, namely
\begin{subequations}
 \begin{align}
  \lim_{\alpha\to\infty} A_n &= -\left( n+\frac{3}{2} \right) \frac{1}{R^{n+1}} + \left( n+\frac{1}{2} \right) \frac{1}{R^{n+3}} \, , \label{A_n_hardSphere} \\
 \lim_{\alpha\to\infty} B_{n} &= -\frac{1}{4}(1-R^2)^2 \frac{1}{R^{n+5}} \, , \label{B_n_hardSphere}
 \end{align}
\end{subequations}
in agreement with the results by Kim and Karrila \cite{kim13} [p.~243].


\paragraph{Bending contribution}

In the following, we consider an idealized membrane with a bending-only resistance such as an artificial vesicle.
By setting $\Delta f_r^{\mathrm{S}} = \Delta f_\theta^{\mathrm{S}} = 0$ in the traction jump equations given by Eqs.~\eqref{BC:sigma_r_phi} and \eqref{BC:sigma_r_r}, we get
\begin{subequations}
 \begin{align}
  \left[ v_{\theta,r} \right] &= \left. \alphaB \left( \left(1-\cot^2\theta\right)v_{r,\theta} + v_{r,\theta\theta} \cot\theta + v_{r,\theta\theta\theta} \right) \right|_{r=1} \, , \label{jump_v_Phi_Bending} \\
  \bigg[ -\frac{p}{\eta} \bigg] &=  \alphaB \bigg( \left( 3\cot\theta+\cot^3\theta \right) v_{r,\theta} -  v_{r,\theta\theta}\cot^2\theta \notag \\
                                            &+ \left. 2 v_{r,\theta\theta\theta}\cot\theta + v_{r,\theta\theta\theta\theta} \bigg) \right|_{r=1} \, , \label{jump_v_R_Bending}
 \end{align}
\end{subequations}
where $i \alphaB := \kB /(\eta \omega)$.
Note that the right hand side of Eq.~\eqref{jump_v_R_Bending} stands for the tangential biharmonic operator \cite{spiegel68} applied to the velocity radial component $v_r$.

We then substitute the expressions of the velocity components given by Eqs.~\eqref{v_r_S}-\eqref{v_phi_Inside} into the tangential traction jump Eq.~\eqref{jump_v_Phi_Bending} and replace $a_n$ and $b_n$ with their expressions given respectively by Eqs.~\eqref{a_n} and \eqref{b_n}.
After multiplying both members by $\psi_m \sin\theta$, preforming the integration between 0 and $\pi$, 
and making use of the orthogonality identities \eqref{orthoProp_psi_1} and \eqref{orthoProp_psi_2} together with Eq.~\eqref{psi_n}, we obtain 
\begin{equation}
 \begin{split}
  &(2n+1) \Big( 2(2n+3) B_{n-1} - A_{n-1} + A_{n+1} \Big) = \\
  &\alphaB  \bigg( \frac{A_{n+1}}{2n+3} -\frac{A_{n-1}}{2n-1}  + 2 B_{n-1}
                               -\frac{1}{2n-1} \frac{1}{R^{n}} \\
                               &+ \frac{1}{2n+3} \frac{1}{R^{n+2}} \bigg) 
                               n(n+1)(-n^2-n+2) \, ,
 \end{split}
\label{jump_f_Phi_Bending_Finalize()}
\end{equation}
for $n\ge 1$.


Next, after substitution in the normal traction jump Eq.~\eqref{jump_v_R_Bending}, multiplying both members by $\varphi_m \sin\theta$ and using Eq.~\eqref{orthoProp_phi_1} together with the orthogonality identity
\begin{equation}
 \begin{split}
  &\int_{0}^{\pi} \varphi_m \bigg( \left( 3\cot\theta+\cot^3\theta \right) \varphi_{n,\theta} -  \varphi_{n,\theta\theta}\cot^2\theta \\
      &+ 2 \varphi_{n,\theta\theta\theta}\cot\theta + \varphi_{n,\theta\theta\theta\theta} \bigg) \sin\theta \Intd \theta \\
      &= \frac{2n(n-1)(n+1)(n+2)}{2n+1} \frac{\delta_{mn}}{r^{2n+2}} \, , \notag
 \end{split}
\end{equation}
we get after replacing $a_n$ and $b_n$ with their corresponding expressions 
\begin{equation}
 \begin{split}
   &-2(2n+3)(2n+1)(n+1)B_{n-1} + (2n^2+7n+3) A_{n-1} \\
   &- (2n^2+3n+1)A_{n+1} = \alphaB    \bigg(  \frac{A_{n+1}}{2n+3} -\frac{A_{n-1}}{2n-1}  
			       + 2 B_{n-1} \\
                               &-\frac{1}{2n-1} \frac{1}{R^{n}} + \frac{1}{2n+3} \frac{1}{R^{n+2}} \bigg)
                               (n-1)n^2(n+1)^2(n+2)  \, ,
 \end{split}
 \label{jump_f_R_Bending_Finalize()}
\end{equation}
for $n \ge 1$.

From Eq.~\eqref{jump_f_Phi_Bending_Finalize()}, $B_{n-1}$ can straightforwardly be expressed in terms of $A_{n-1}$ and $A_{n+1}$.
We obtain
\begin{equation}
  \begin{split}
   B_{n} &= -\frac{A_{n+2}}{4n+10} + \frac{1}{S}
  \bigg(
  \frac{S' A_n}{2n+1}  
  +\alphaB n(n+1)(n+2)(n+3) \\
  &\times \left( \frac{1}{2n+1}\frac{1}{R^{n+1}} - \frac{1}{2n+5} \frac{1}{R^{n+3}} \right)
  \bigg) \, , 
  \end{split}
  \label{B_n_Bending}
\end{equation}
for $n \ge 0$,
where we defined
\begin{subequations}
  \begin{align}
   S  &:= 2\Big( \alphaB n^4+6\alphaB n^3+(11\alphaB+4)n^2  
      + (6\alphaB+16)n+15 \Big) \, , \notag \\
   S' &:= S/2 - 8n-12 \, . \notag 
  \end{align}
\end{subequations}

After plugging the expression of $B_{n-1}$ into Eq.~\eqref{jump_f_R_Bending_Finalize()}, we get the general term of $A_n$ as
\begin{equation}
 A_n =  \frac{\alphaB n^2(n+1)(n+3)(n+2)^2}{W} \left( \frac{2n+1}{R^{n+3}} - \frac{2n+5}{R^{n+1}} \right) \, , 
 \label{A_n_Bending}
\end{equation}
for $n \ge 0$, where 
\begin{equation}
  \begin{split}
   W &:= 4\alphaB n^6+36\alphaB n^5+118\alphaB n^4+(168\alphaB+16)n^3 \notag \\
    &+(94\alphaB+72)n^2+(12\alphaB+92)n+30 \, . \notag 
  \end{split}
\end{equation}

The general term for $B_n$ can be obtained by substituting $A_{n}$ and $A_{n+2}$ as computed from Eq.~\eqref{A_n_Bending} into Eq.~\eqref{B_n_Bending}.
Interestingly, by taking $\alphaB$ to infinity, $A_n$ and $B_n$ do not tend to the hard-sphere limits as it has been shown to be the case for a shearing-only membrane.
In this case we rather obtain
\begin{subequations}
 \begin{align}
  \lim_{\alphaB\to\infty} A_n &= \frac{n(n+2)}{2(2n^2+6n+1)} \left( \frac{2n+1}{R^{n+3}} - \frac{2n+5}{R^{n+1}} \right) \, , \notag \\
  \lim_{\alphaB\to\infty} B_n &= \frac{1}{4} \bigg( -\frac{n^2+2n-2}{2n^2+6n+1} - \frac{(n+2)(n+4)}{2n^2+14n+21} \frac{1}{R^4} \notag \\
                              &+\frac{2n^4+18n^3+49n^2+42n+3}{(2n^2+6n+1)(2n^2+14n+21)} \frac{2}{R^2} \bigg) \frac{1}{R^{n+1}} \, . \notag
 \end{align}
\end{subequations}

A similar resolution approach can be adopted for the determination of the series coefficients when the membrane is simultaneously endowed with both shearing and bending rigidities.
Analytical expression can be obtained by computer algebra software, but they are not included here due to their complexity and lengthiness.
We note that the shearing and bending contributions to the particle mobility do not superpose linearly which is in contrast to a planar membrane \cite{daddi16} but similar to what has been observed between two planar membranes \cite{daddi16b}.


\section{Particle self-mobility}\label{sec:particleMobility}

In this section, we compute the correction to the particle self-mobility in the point-particle framework.
Here we assume no net force on the capsule and an external force $\Fext$ on the solid particle.
As shown in Appendix~\ref{appendix:forceFree}, for finite membrane shearing modulus, the capsule is in fact force free.

The zeroth-order solution for the particle velocity is given by the Stokes law as $\vect{V}_2^{(0)} = \mu_0 \Fext$, where $\mu_0 := 1/(6\pi\eta b)$ is the usual bulk mobility.
The first-order correction to the particle self-mobility $\Delta\mu$ is obtained by evaluating the reflected flow field at the particle position such that 
\begin{equation}
    \vect{v}^*|_{\vect{x}=\xTwo} = \Delta\mu \Fext \, .
\end{equation}
Since the force points along the axis of symmetry of the system, the mobility correction is a simple scalar and not a tensor as it would be for an arbitrary direction of the force.
In the following, we shall make use of the following identities 
\begin{subequations}
 \begin{align}
  \left. \frac{(\vecD\cdot\boldNabla)^n}{n!} \Gmatr (\R) \right|_{\vect{x}=\xTwo}  \Fext &= \frac{2}{R^{n+1}} \Fext \, , \notag \\
  \left. \frac{(\vecD\cdot\boldNabla)^n}{n!}  \boldNabla^2 \Gmatr (\R) \right|_{\vect{x}=\xTwo} \Fext &= -\frac{2(n+1)(n+2)}{R^{n+3}} \Fext \, . \notag
 \end{align}
\end{subequations}
to finally obtain
\begin{equation}
  \frac{\Delta \mu}{\mu_0} = \frac{3b}{4} \infSum 2\left( A_n-(n+1)(n+2)\xi^2 B_n \right) {\xi^{n+1}}  \, ,
	    \label{mobilityCorrection}
\end{equation}
wherein $\xi:=1/R \in [0,1)$.
This is the central result of our work.
We recall that the unscaled form for an arbitrary capsule radius $a$ can be obtained from Eq.~\eqref{mobilityCorrection} by the replacement rules in Appendix~\ref{appendix:transformationEquations}.
The number of terms to be included before the series is truncated can be estimated for a desired precision as detailed in appendix~\ref{appendix:seriesEstimation}. 
{Due to the point-particle approximation, the particle radius only enters upon rescaling the particle self-mobility correction by the bulk mobility~$\mu_0$.}

\subsection{Shearing contribution}

For a membrane exhibiting a shearing-only resistance, the particle self-mobility correction can be computed by plugging the expressions of $B_n$ and $A_n$ as stated respectively by Eqs.~\eqref{B_n_Shearing} and \eqref{A_n_Shearing} into Eq.~\eqref{mobilityCorrection}.
By taking the limit when $\alpha\to\infty$ we recover the rigid sphere limit, 
\begin{equation}
 \frac{\Delta \mu_{\mathrm{S}, \infty}}{\mu_0} := \lim_{\alpha\to\infty}  \frac{\Delta \mu_\mathrm{S}}{\mu_0} = -\frac{\xi^3 (15-7\xi^2+\xi^4)}{4(1-\xi^2)} \frac{b}{R} \, ,
 \label{hardSphere}
\end{equation}
in agreement with the result by Ekiel-Je{\.z}ewska and Felderhof \cite[Eq. (2.26)]{ekiel15}.
For an infinite membrane radius,  we obtain
\begin{equation}
  \frac{\Delta \mu_{\mathrm{S}, \infty}}{\mu_0}  = -\frac{9}{8} \frac{b}{h}  \, , \label{hardWall_series}
\end{equation}
where $h := R-1$ being the distance from the center of the solid particle to the closest point on the capsule surface.
We thus recover the well-known result for a planar rigid wall as first calculated by Lorentz about one century ago \cite{lorentz07}.


\begin{figure}
 \includegraphics[width=0.5\textwidth]{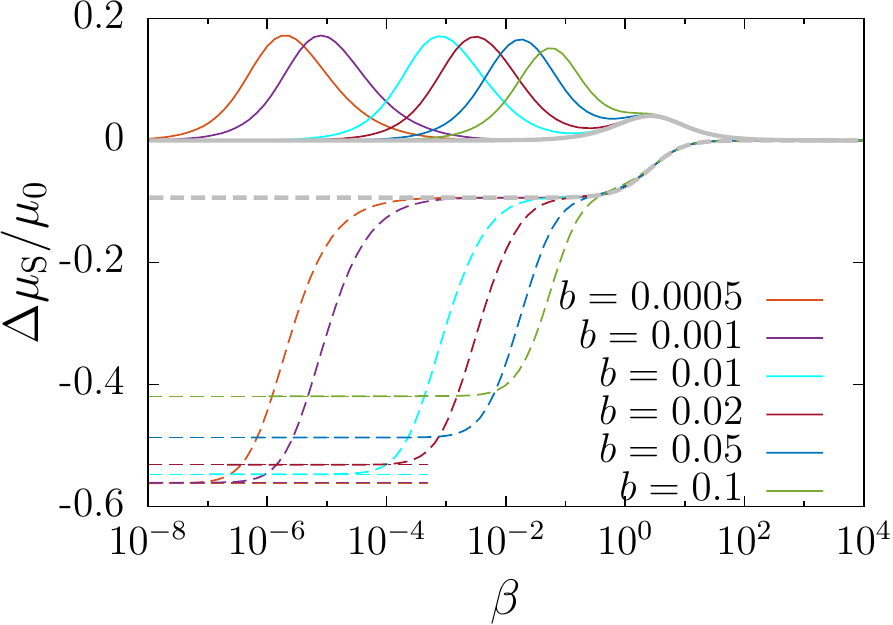}
 \caption{(Color online) Scaled particle self-mobility correction versus  $\beta$ for various values of $b$ for a shearing-only membrane. 
 The real and imaginary parts are shown as dashed and solid lines respectively.
 Horizontal dashed lines represent the hard-sphere limit as given by Eq.~\eqref{hardSphere}.
 The curve in gray corresponds to the self-mobility correction for a planar membrane as given by Eq.~\eqref{delta_Mu_Shearing_Planar}.
 Here we set the solid particle at $h=2b$.}
 \label{sphericalPerpShearing_A_Effect}
\end{figure}

\begin{figure}
 \includegraphics{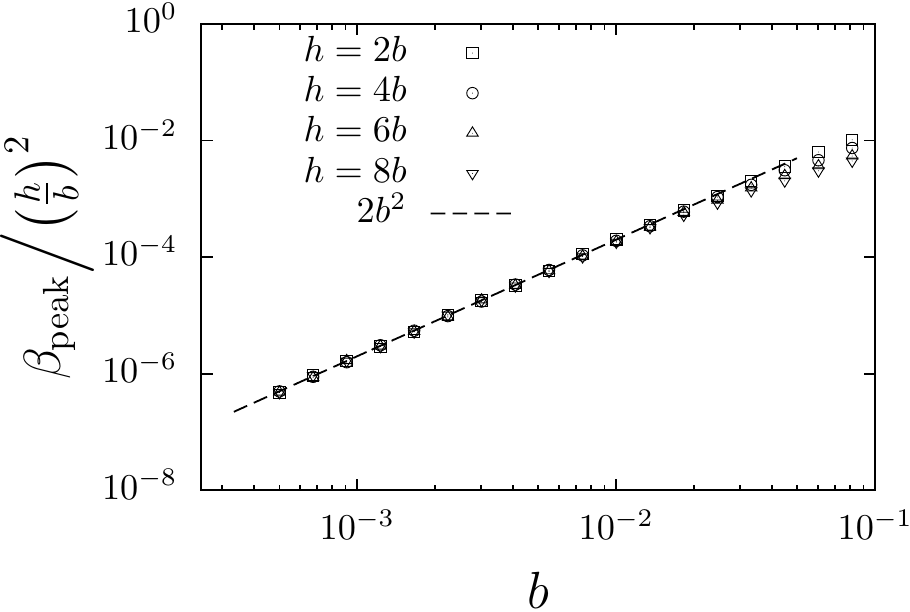}
 \caption{ Log-log plot of the rescaled peak-frequency versus particle radius for different particle-to-membrane distance $h$.
 }
 \label{betaPeak}
\end{figure}

We define the characteristic frequency for shearing as $\beta := 6B\eta\omega h/\kS$ with $B:=2/(1+C)$.
In Fig.~\ref{sphericalPerpShearing_A_Effect} we plot the variations of the scaled self-mobility correction for a shearing-only membrane versus $\beta$ upon varying the particle radius $b$ while keeping the distance from the membrane $h=2b$ and setting the Skalak parameter $C=1$. 
We observe that the real part of the mobility correction is a monotonically increasing function of frequency and the imaginary part exhibits the typical peak structure which is a signature of the memory effect induced by the elastic nature of the membrane.
In the vanishing frequency limit, the correction is identical to that near a hard-sphere with stick boundary conditions, given by Eq.~\eqref{hardSphere}.

For sufficiently small values of $b$ (or equivalently for larger capsule radii), we observe that in the high frequency regime for which $\beta \geq 1$, both the real and imaginary parts of the mobility correction follow faithfully the evolution of those predicted for a planar membrane which is \cite{daddi16}
\begin{equation}
 \frac{\Delta\mu_\mathrm{S} (\beta)}{\mu_0} = -\frac{9}{16} \frac{b}{h} e^{i\beta} \E_4(i\beta) \, .
 \label{delta_Mu_Shearing_Planar}
\end{equation}
{The peak position around $\beta\sim 1$ can be estimated by a simple balance between membrane elasticity and fluid viscosity as $\omega \sim \kappa_\mathrm{S}/(\eta h)$.}
{A strong departure} is however observed in the low frequency regime where a second peak of more pronounced amplitude occurs in the imaginary part.
This second peak is the most prominent signature which distinguishes the spherical membrane from the planar case.
{The peak height remains typically constant for a large range of values of $b$ because the mobility correction has been rescaled by the bulk mobility.}

{We attribute the two peaks in Fig.~2 to in-plane deformations ($u_\theta$) and surface-normal deformations ($u_r$), respectively. 
The radius-independent peak around $\beta\sim 1$ corresponds to in-plane deformations $u_\theta$ which are present in a similar way for the planar membrane thus explaining the agreement with Eq.~\eqref{delta_Mu_Shearing_Planar}.
The larger and radius-dependent peak corresponds to surface-normal deformations which contribute to the traction jump even for a shear-only membrane as can be seen in Eq.~\eqref{tractionJumpShear}.
This contribution is due to the membrane curvature: in the planar case, surface-normal deformations do not contribute to the traction jump associated with shear at first order (cf. equation (A20) of Ref.~\onlinecite{daddi16}) and therefore this peak is not observed for the planar membrane.
Indeed, upon increasing the capsule radius (decreasing $b$), the second peak gradually shifts towards lower frequencies and eventually disappears for $b\to 0$.
}

In Fig.~\ref{betaPeak}, we plot the variations of the rescaled peak frequency occurring in the imaginary part of the particle self-mobility versus particle radius $b$ at different values of~$h$.
{For sufficiently small particles ($b < 0.05$), the peak frequency shows a quadratic increase with particle radius~$b$.
By rescaling the peak frequencies by $(h/b)^2$, a master curve is obtained and the peak frequency position can accurately be computed from the relation $\beta_\mathrm{peak} = 2 h^2$.}


\subsection{Bending contribution}

For a bending-only membrane, the mobility correction is readily obtained after plugging the series coefficients $B_n$ and $A_n$ respectively given by Eqs.~\eqref{B_n_Bending} and \eqref{A_n_Bending} into Eq.~\eqref{mobilityCorrection}.
In particular, by taking $\alphaB \to \infty$, the leading order self-mobility correction can conveniently be approximated by
\begin{equation}
 \frac{\Delta\mu_{\mathrm{B}, \infty}}{\mu_0}:= \lim_{\alphaB \to \infty}  \frac{\Delta\mu_\mathrm{B} }{\mu_0} \simeq -\frac{7\xi^3}{4(1-\xi^2)} \left[1 + \frac{\xi^2}{5}  - \frac{9 \xi^4}{70}  \right] \frac{b}{R} \, , 
 \label{vanishing_Frequency_Bending}
\end{equation}
which, for an infinite radius reads
\begin{equation}
   \frac{\Delta\mu_{\mathrm{B}, \infty}}{\mu_0} = -\frac{15}{16} \frac{b}{h}  \, , \label{bendingLimit}
\end{equation}
corresponding to the vanishing frequency limit for a planar membrane with bending-only as calculated in earlier work \cite{daddi16}.
Note that this limit is the same as that for { a flat fluid-fluid interface separating two immiscible liquids having the same dynamic viscosity \cite{lee79}.}

\begin{figure}
 \includegraphics{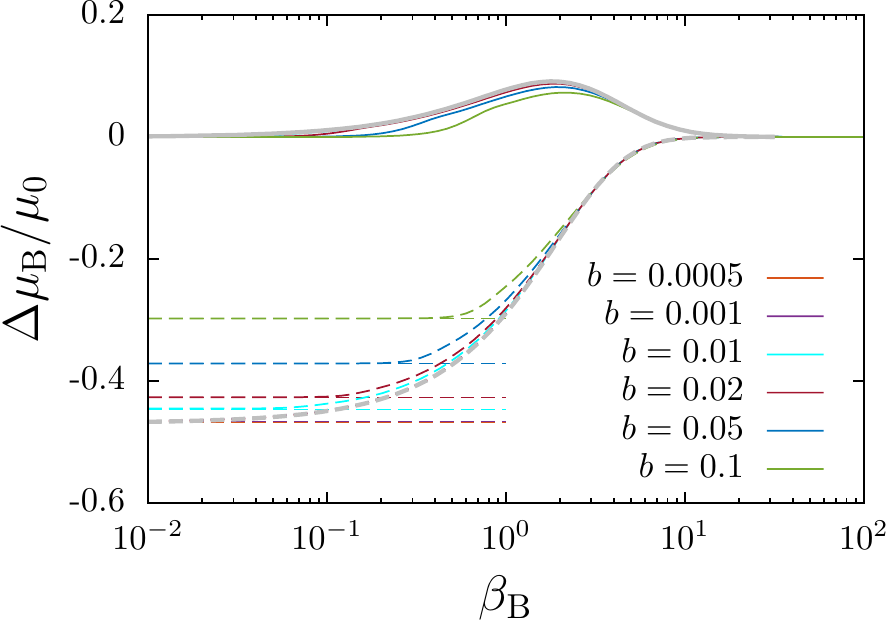}
 \caption{(Color online) Scaled self-mobility correction versus $\betaB$ for various values of the capsule radius, for a bending-only membrane.
 The dashed and continuous lines represent the real and imaginary parts respectively.
 The horizontal dashed lines are the vanishing frequency limits as approximated be Eq.~\eqref{vanishing_Frequency_Bending}.
 The curve in gray is the solution for a planar membrane given by Eq.~\eqref{delta_Mu_Bending_Planar}.
 Here we take $h=2b$.
 }
 \label{sphericalPerpBending_A_Effect}
\end{figure}

We define the characteristic frequency for bending as $\betaB :=  2h(4\eta \omega / \kB)^{1/3} $.
In Fig.~\ref{sphericalPerpBending_A_Effect}, we present the particle self-mobility correction nearby a membrane exhibiting a bending-only resistance versus $\betaB$.
{Unlike a membrane with shearing resistance only, the particle mobility correction nearby a bending-only membrane does not exhibit a second peak of pronounced amplitude.
The single peak observed is the characteristic peak for bending which occurs at $\beta_\mathrm{B}^3 \sim 1$ and is largely independent of the radius.
In fact, this peak position can be estimated by a balance between fluid viscosity and membrane bending such that $\omega \sim \kappa_\mathrm{B}/(\eta h^3)$.
As can be seen from equations~\eqref{tractionJumpBend}, the traction jump for a bending-only membrane involves only the radial deformation which explain the absence of a second peak in contrast to the two-peak structure seen in the shearing-only case.
}

As already pointed out {in Sec.~\ref{sec:singularitySolution},} the hard-sphere solution is not recovered for a bending-only membrane in the vanishing frequency limit.
{ 
A similar trend has been observed in earlier work for planar membranes where bending alone is not sufficient to recover the hard-wall limit \cite{daddi16}.
This feature is again justified by the fact that the traction jumps due to bending in Eq.~\eqref{tractionJumpBend} do not depend on the tangential displacement $u_\theta$.
Even when considering an infinite bending modulus $\kappa_\mathrm{B}$, the tangential component of the membrane displacement is thus still completely free.
This behavior cannot represent the hard sphere where both radial and tangential displacements are restricted.
}

We further remark that for smaller values of $b$, the evolution of both the real and imaginary part is found to be in excellent agreement with the solution   for a planar membrane \cite{daddi16} in the whole range of frequencies:
\begin{equation}
 \begin{split}
  \frac{\Delta\mu_\mathrm{B} (\betaB)}{\mu_0} &= \frac{3i\betaB}{8} \frac{b}{h} \bigg( 
			  \left( \frac{\betaB^2}{12}+\frac{i\betaB}{6}+\frac{1}{6} \right) \phi_{+} \\
			  &+\frac{\sqrt{3}}{6}(\betaB+i)\phi_{-} + \frac{5i}{2\betaB} \\
			  &+ \left( \frac{\betaB^2}{12} - \frac{i\betaB}{3} - \frac{1}{3} \right) e^{-i\betaB} \E_1(-i\betaB)
			  \bigg) \, , 
 \end{split}
 \label{delta_Mu_Bending_Planar}
\end{equation}
with
\begin{equation}
 \phi_{\pm} = e^{-i\overline{\zB}} \E_1(-i\overline{\zB}) \pm e^{-i\zB} \E_1(-i\zB) \, , 
\end{equation}
where $\zB := \betaB e^{2i\pi/3}$.

We therefore conclude that for large capsules, the mobility correction for a bending-only membrane can be appropriately estimated from the planar membrane limit.
For moderate capsule radii, the planar membrane prediction gives a reasonable agreement only in the high frequency regime for which $\betaB \gg 1$.

\subsection{Shearing-bending coupling}

\begin{figure}
 \includegraphics{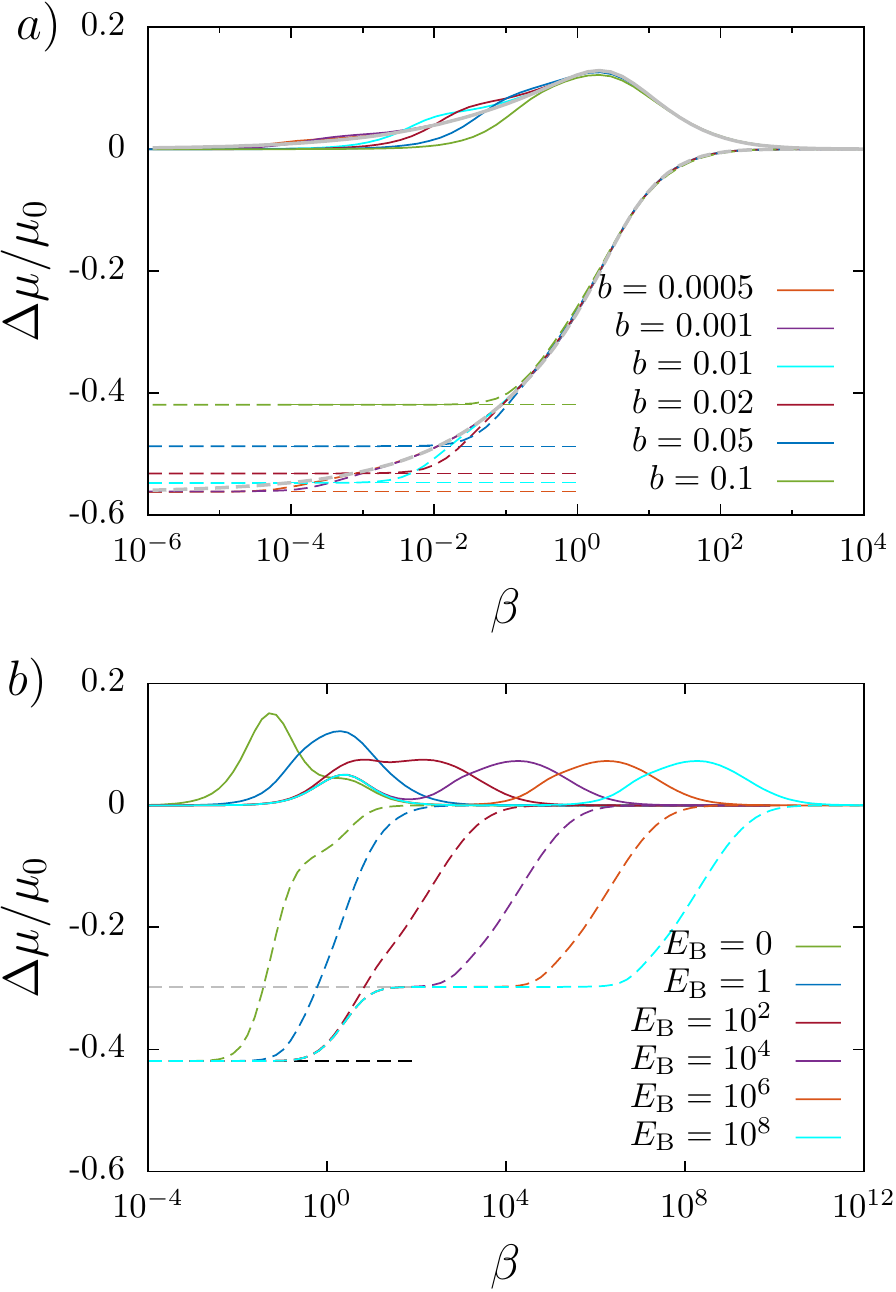}
 \caption{(Color online) $a)$ Scaled particle self-mobility correction versus $\beta$ for various values of the particle radius $b$ for a membrane endowed with both shearing and bending rigidities. 
 The real and imaginary parts are shown as dashed and solid lines respectively.
 Horizontal dashed lines represent the hard-sphere limit as given by Eq.~\eqref{hardSphere}.
 The curve in gray corresponds to the self-mobility correction for a planar membrane as obtained by linear superposition of Eqs.~\eqref{delta_Mu_Shearing_Planar} and \eqref{delta_Mu_Bending_Planar}.
 Here we set the solid particle at $h=2b$ and we take a reduced bending modulus $\EB=1$.
 $b)$ Scaled self-mobility correction versus $\beta$ for various values of the reduced bending modulus.
 The horizontal black dashed line is the hard-sphere limit given by Eq.~\eqref{hardSphere} whereas the gray dashed line is the infinite bending rigidity limit for a bending-only membrane as given by Eq.~\eqref{vanishing_Frequency_Bending}.
 Here we take $b=1/10$ and  $h=2b$.
 }
 \label{Fig_EB}
\end{figure}

{
Unlike for a single planar membrane, shearing and bending are intrinsically coupled for a spherical membrane and the particle mobility near a membrane exhibiting shearing and bending resistance cannot be obtained by linear superposition as for a planar membrane \cite{daddi16}.
A similar coupling is also observed for the mobility of a particle between two parallel planar membranes \cite{daddi16b} as well as for thermal fluctuations of two closely-coupled 
\cite{auth07} or ''warped'' \cite{kovsmrlj14} membranes.
Therefore, the solution requires to simultaneously consider shearing and bending in the traction jump equations.
}
In order to investigate this coupling effect, we define the reduced bending modulus as $\EB:=\kB/(\kS h^2)$, a parameter that quantifies the relative contributions of shearing and bending.

In Fig.~\ref{Fig_EB}~$a)$ we show the scaled self-mobility correction versus $\beta$ nearby a membrane with both shearing and bending resistances upon varying $b$.
We observe that in the high frequency regime, i.e. for $\beta > 1$, the mobility correction follows faithfully the evolution predicted for a planar membrane.
For lower values of $b$, the planar membrane solution provides a very good estimation even deeper into the low frequency regime.
Here, we take $h=2b$ and a reduced bending modulus $\EB=1$, {for which shearing and bending manifest themselves equally.}

In Fig.~\ref{Fig_EB}~$b)$, we show the mobility correction versus $\beta$ for a membrane with both rigidities upon varying the reduced bending modulus $\EB$ while keeping $b=1/10$ and $h=2b$.
{For $E_\mathrm{B} = 0$ corresponding to a shearing-only membrane, a low frequency peak as in Fig.~\ref{sphericalPerpShearing_A_Effect} is observed.
For $E_\mathrm{B} \approx 1$ and above, this peak quickly disappears which confirms our hypothesis that it is due to radial deformations as reasoned above:
In the case of large bending resistance these deformations are suppressed and therefore the peak height diminishes and eventually disappears.}

The imaginary part exhibits an additional peak {of typically constant height} that is shifted progressively toward the higher frequency domain {for increasing $\EB$.}
{From the definitions of $\beta$ and $\beta_\mathrm{B}$, it can be seen that
\begin{equation}
 \beta_\mathrm{B}^3 = \frac{16}{3B E_\mathrm{B}} \, \beta \, .
\end{equation}
Therefore, the peak observed at $\beta \sim 1$ is attributed to shearing whereas the high frequency peak is attributed to bending because $\beta \sim E_\mathrm{B}$ when $\beta_\mathrm{B}^3 \sim 1$.
Particularly, for $E_\mathrm{B} = 1$, the  position of the two peaks coincides as $\beta \sim \beta_\mathrm{B}^3$ for which shearing and bending have equal contribution.}

\section{Capsule motion and deformation}\label{sec:capsuleMotion}

Next, we examine the capsule motion induced by the nearby moving solid particle.
For this aim, we define the pair-mobility function $\muP$ as the ratio between the centroid velocity of the capsule $V_1$ and the force $F_2$ applied on the solid particle, i.e. $V_1 = \muP F_2$.
The net translational velocity of the capsule can readily be computed by volume integration of the $z$ component of the fluid velocity inside the capsule  \cite{felderhof06b},
\begin{equation}
 V_1 (\omega) = \frac{2\pi}{\Omega} \int_0^{\pi} \int_0^1 v_z^{(i)} (r,\theta, \omega) \, r^2 \sin \theta \,  \Intd r \Intd \theta \, , 
\end{equation}
where $\Omega := 4\pi/3$ being the volume of the undeformed capsule and $v_z^{(i)} = v_r^{(i)} \cos \theta - v_{\theta}^{(i)} \sin \theta$.
Analytical expressions for $v_r^{(i)}$ and $v_\theta^{(i)}$ are given by Eqs.~\eqref{v_r_Inside} and \eqref{v_phi_Inside} respectively.
After computation, only the terms with $n=1$ of the series remain.
The  frequency-dependent pair-mobility reads
\begin{equation}
  {\muP}  = - \frac{1}{8\pi\eta} (a_1 + b_1) \, , 
\end{equation}
which can be simplified to obtain
\begin{equation}
 6\pi \eta {\muP}  = \frac{3}{2}\xi -\frac{\xi^3}{2} \frac{3+(1+2C) \alpha}{5+(1+2C) \alpha} \, . \label{pair-mobi}
\end{equation}

The leading order pair-mobility correction is therefore expressed as a Debye-type model with a relaxation time given by
\begin{equation}
 \tau = \frac{15}{2(1+2C)} \frac{\eta}{\kS} \, .
\end{equation}

Interestingly, the pair-mobility $\muP$ depends only on the shear resistance of the membrane, but not on membrane bending properties.
In the limiting cases, we recover two known results.
First, for an infinite membrane shearing modulus, we get the leading-order pair-mobility between two unequal hard-spheres
\begin{equation}
 \lim_{\alpha \to \infty} 6\pi \eta {\muP}  = \frac{3}{2} \xi - \frac{\xi^3}{2} \, . \label{pair-mobi-hard-sphere}
\end{equation}
Second, for a vanishing membrane shearing modulus, we obtain the leading-order pair-mobility between a solid particle and a viscous drop 
\begin{equation}
 \lim_{\alpha \to 0} 6\pi \eta {\muP} = \frac{3}{2} \xi - \frac{3}{10} \xi^3 \, , \label{pair-mobi-drop}
\end{equation}
both of which are in agreement with those reported by Fuentes \textit{et al.} \cite[Eq. (12)]{fuentes88}.

\subsection*{Membrane deformation}

In this subsection, we compute the capsule deformation resulting from an arbitrary time-dependent point-force $F$ acting nearby the spherical capsule.
The membrane displacement field is related to the fluid velocity at $r=1$ via the no-slip equation given by Eq.~\eqref{no-slip}.
In order to proceed, we define the frequency-dependent reaction tensor $\psi_{ij}$ as
\begin{equation}
 u_{i} (\theta, \omega) = \psi_{ij} (\theta, \omega) F_j (\omega) \, . \label{reactionTensorDef}
\end{equation}

By setting a harmonic driving force $F_i (t) = K_i e^{i\omega_0 t}$, which in the frequency domain reads $F_i (\omega) = 2\pi K_i \delta(\omega-\omega_0)$, 
the membrane time-dependent displacement can readily be evaluated by inverse Fourier transform of Eq.~\eqref{reactionTensorDef} to obtain
\begin{equation}
 u_i (\theta, t) = \psi_{ij} (\theta, \omega_0) K_j e^{i\omega_0 t} \, .
\end{equation}

In an axisymmetric situation, we are interested in the components $\psi_{rz}$ and $\psi_{\theta z}$ of the reaction tensor, giving access to the displacements $u_r$ and $u_\theta$ under the action of a point force directed along the $z$ direction.
By making use of Eqs.~\eqref{v_r_Inside} and \eqref{v_phi_Inside}, we immediately obtain
\begin{subequations}
 \begin{align}
  \psi_{rz}  &= -\frac{1}{8\pi\eta i\omega} \sum_{n=1}^{\infty} \left[ \frac{n(n+1)}{2} a_n + n b_n \right] P_n(\cos \theta) \, , \label{psi_rz} \\
  \psi_{\theta z} &=  -\frac{1}{8\pi\eta i \omega} \sum_{n=1}^{\infty} \left[ \frac{n+3}{2} a_n + b_n \right] \frac{\Intd P_n(\cos \theta)}{\Intd \theta} \, . \label{psi_phiz}
 \end{align}
\end{subequations}
The first derivative of Legendre polynomial can be computed using the recurrence formula \cite{abramowitz72}
\begin{equation}
 \frac{\Intd P_n(\cos\theta)}{\Intd \theta} = -\frac{n}{\sin\theta} \Big[ P_{n-1} (\cos\theta) - \cos\theta P_{n}(\cos\theta) \Big] \, . \notag
\end{equation}

\section{Comparison with boundary integral simulations}\label{sec:comparisonWithBIM}

\begin{figure}
 \includegraphics{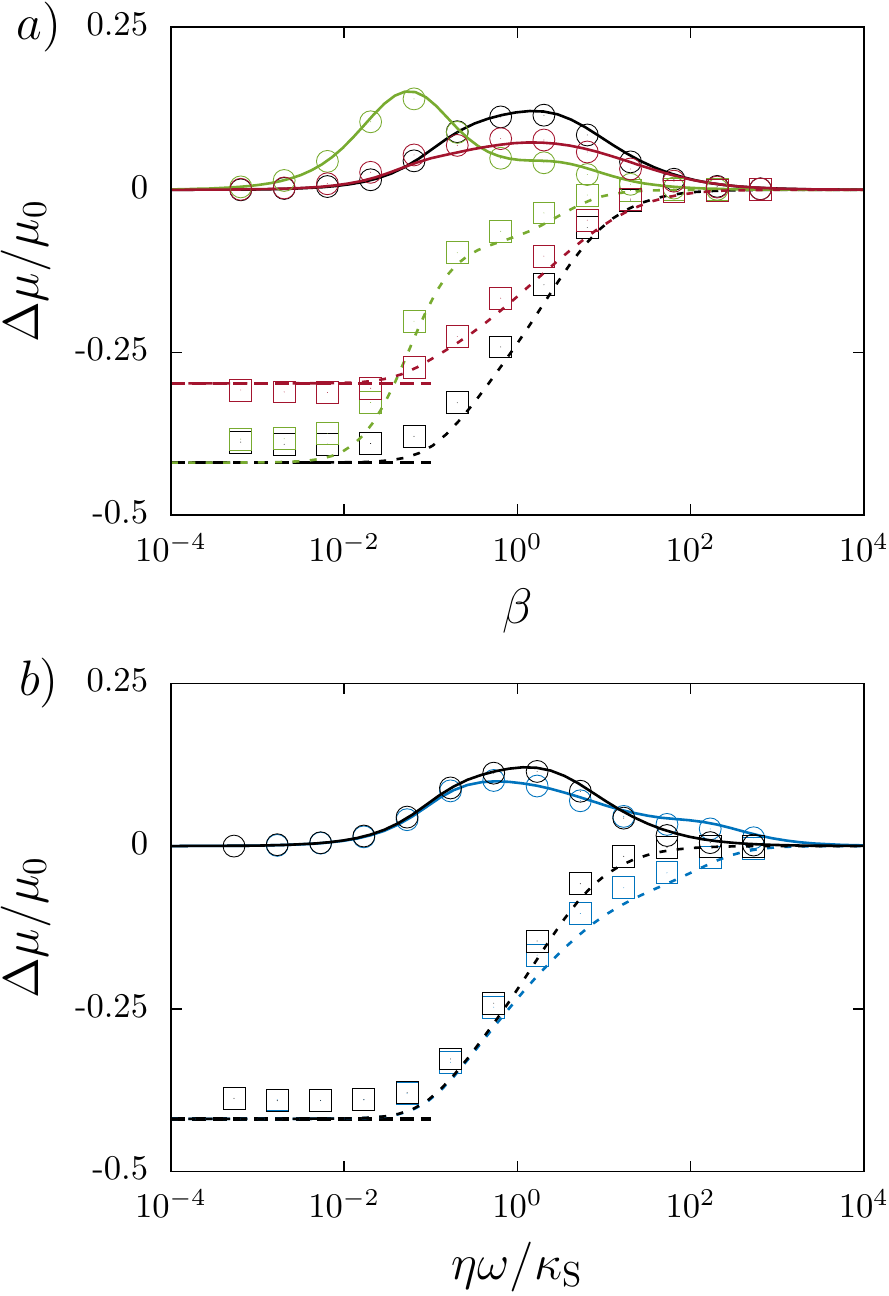}
 \caption{(Color online) $a)$ Scaled frequency-dependent particle mobility correction versus the scaled frequency $\beta$ nearby a membrane endowed with only shearing (green / light gray), only bending (red / dark gray) and both rigidities (black).
 The small particle has a radius $b=1/10$ set a distance $h=2b$.
 Here we take $C=1$ and a reduced bending modulus $\EB=2/3$.
 The theoretical predictions are shown as dashed lines for the real parts and as solid lines for the imaginary parts.
 Symbols refer to boundary integral simulations where the real and imaginary parts are shown as squares and circles respectively.
 The horizontal dashed lines are the vanishing frequency limits given by Eqs.~\eqref{hardSphere} and \eqref{vanishing_Frequency_Bending}.
 $b)$ shows the scaled frequency-dependent mobility correction versus $\eta\omega/\kS$ nearby a membrane endowed with  both shearing and bending rigidities for $C=1$ (black) and $C=100$ (blue / dark gray) for the same set of parameters in $a)$.
 }
 \label{anaNum}
\end{figure}

\begin{figure}
 \includegraphics{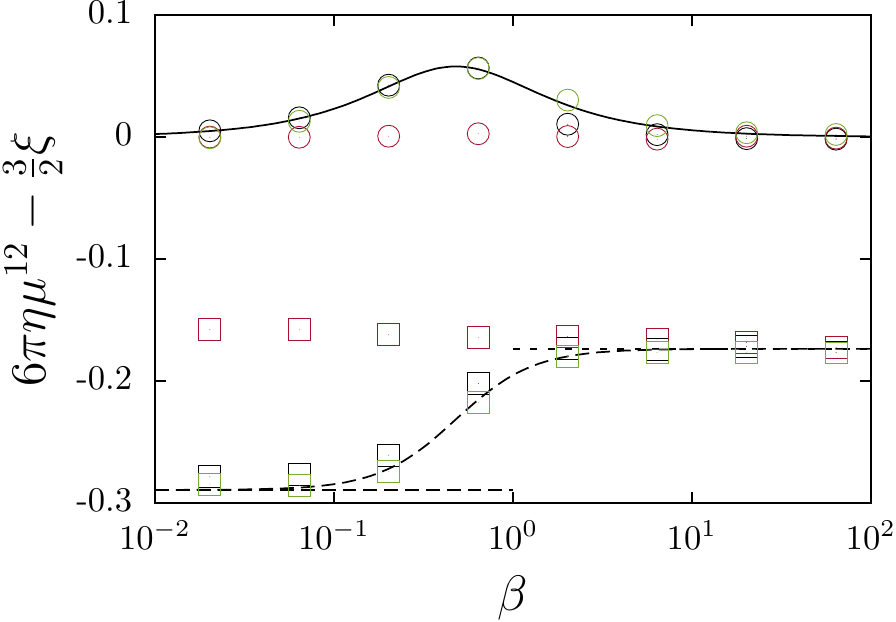}
 \caption{ (Color online) Scaled Pair-mobility correction versus the scaled frequency nearby a membrane possessing only shearing (green / light gray), only bending (red / dark gray) and both rigidities (black). 
 The analytical prediction given by Eq.~\eqref{pair-mobi} is shown as dashed line for the real part and as solid line for the imaginary part.
 Simulation results are shown as squares and circles for the real and imaginary parts, respectively.
 The horizontal dashed lines are the vanishing frequency limit predicted by Eq.~\eqref{pair-mobi-hard-sphere} where the dotted lines are the limit corresponding to vanishing membrane moduli as given by Eq.~\eqref{pair-mobi-drop}.
 }
 \label{capsPairMobi}
\end{figure}

\begin{figure}
 \includegraphics{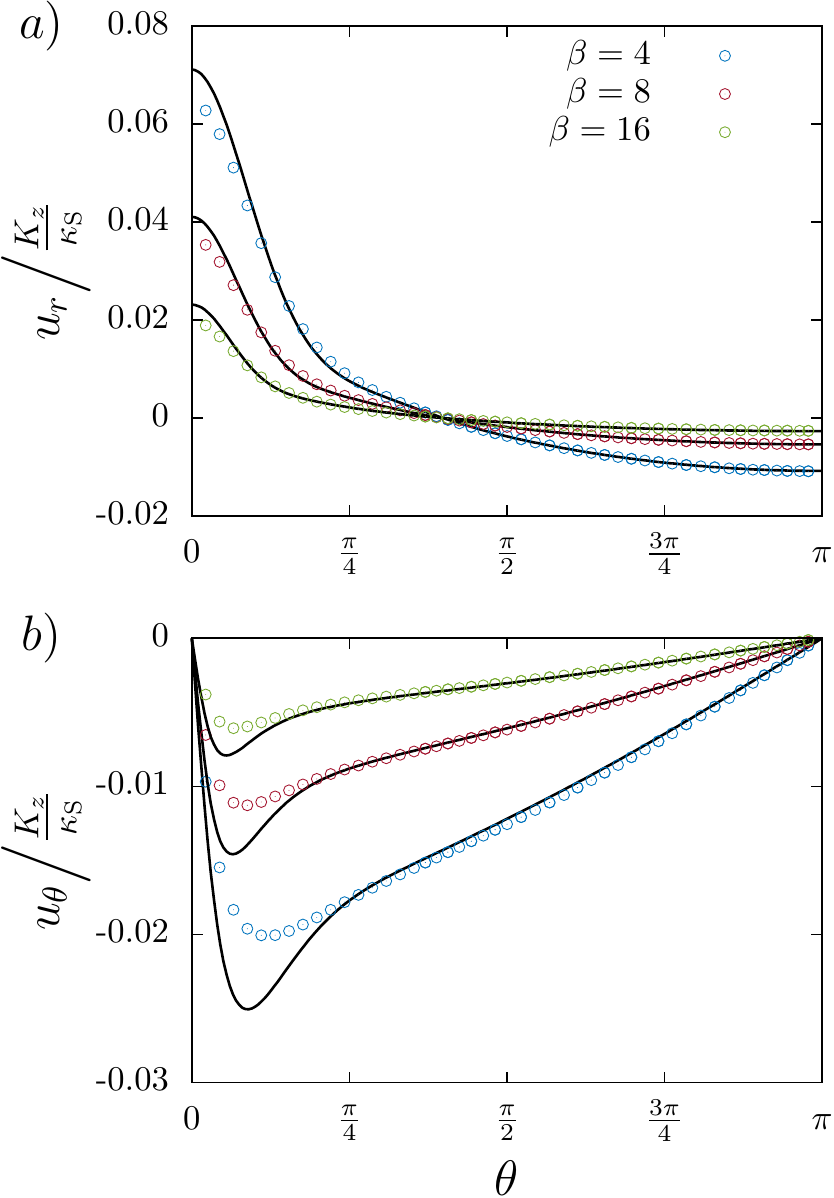}
 \caption{(Color online) Scaled radial $a)$ and meridional $b)$ membrane displacement versus the polar angle $\theta$ for three scaled forcing frequencies $\beta$ at quarter period for $t\omega_0 = \pi/2$.
 Solid lines are the theoretical predictions obtained from Eqs.~\eqref{psi_rz} and \eqref{psi_phiz} and symbols are boundary integral simulations.
 }
 \label{deformation}
\end{figure}

In order to assess the appropriateness of the point particle approximation employed throughout this work, we shall compare our analytical predictions with fully resolved boundary integral simulations of truly extended particles.
The simulations are based on the completed double-layer boundary integral equation method (CDLBIEM) \cite{kohr04, zhao11, zhao12} which allows for the efficient simulation of deformable as well as truly solid objects.
Details on the algorithm and its implementation have been reported elsewhere, see for instance Refs.~\onlinecite{daddi16b, guckenberger16, guckenberger16b}.

For the determination of the solid particle self-mobility, a harmonic oscillating force $K e^{i\omega_0 t}$ is applied at the surface of the particle along the $z$ direction.
After a transient evolution, the particle begins to oscillate with the same frequency $\omega_0$ as $V_2 e^{i(\omega_0 t  + \delta_2)}$.
The velocity amplitude $V_2$ and phase shift $\delta_2$ are accurately determined by fitting the numerically recorded velocity using the trust region method \cite{conn00}.
The frequency-dependent self-mobility of the solid particle is then computed as $\mu  = (V_2/K) e^{i \delta_2}$.
Under the effect of the oscillating force, the volume centroid of the capsule undergoes an oscillatory motion along the $z$ direction as $X_1 e^{i(\omega_0 t  + \delta_1)}$.
The capsule pair-mobility is therefore computed as $\muP  = (i\omega_0 X_1/K) e^{i \delta_1}$.


In Fig.~\ref{anaNum}~$a)$, we present the scaled self-mobility correction versus the scaled frequency $\beta$ as given theoretically by Eq.~\eqref{mobilityCorrection}.
The solid particle has a radius $b=1/10$ positioned at $h=2b$ nearby a large capsule.
For the simulations parameters, we take $C=1$ and $\EB=2/3$. 
Results for shearing-only and bending-only membrane are also shown in green and red respectively.
We observe that in the low-frequency regime, the near hard-sphere mobility correction is approached only if the membrane exhibits a resistance towards shearing, in agreement with theoretical calculations.

{
In Fig.~\ref{anaNum}~$b)$, we show the scaled self-mobility correction for $C=1$ and $C=100$.
A very large $C$ is typical for vesicles or red blood cells \cite{krueger11, freund13, gekle16} where the surface area remains almost unchanged during deformation.
We observe that the effect of area expansion is more pronounced in the high frequency regime.
A very good agreement is obtained between analytical predictions and boundary integral simulations over the whole range of applied frequencies.
}

We now turn to the motion of the capsule.
In Fig.~\ref{capsPairMobi}, we show the correction to the scaled pair-mobility versus the scaled frequency $\beta$.
The correction for a shearing-only membrane is almost indistinguishable from that of a membrane with both shearing and bending rigidities.
In the low-frequency regime for which $\beta \ll 1$, the pair-mobility correction approaches that near a hard-sphere given by Eq.~\eqref{pair-mobi-hard-sphere}.
On the other hand, in the high-frequency regime for which $\beta \gg 1$, the correction approaches that near a viscous drop as given by Eq.~\eqref{pair-mobi-drop}.
Moreover, the correction nearby a bending-only membrane remains typically unchanged over the whole range of frequencies, and equals that for a viscous drop.
Indeed, these observations are in complete agreement with the analytical prediction stated by Eq.~\eqref{pair-mobi}.

In Fig.~\ref{deformation}, we show the  membrane scaled radial and meridional displacements versus the polar angle $\theta$ at quarter period for $t\omega_0 = \pi/2$.
{The natural scale for membrane deformation is $K_z/\kS$ the ratio between the forcing amplitude $K_z$ and the shearing resistance $\kS$.}
We observe that the radial displacement $u_r$ is a monotonically decreasing function of $\theta$ and eventually changes sign at some intermediate angle.
On the other hand, the meridional displacement $u_\theta$ is negatively valued and vanishes at $\theta=0$ and $\theta=\pi$ due to the system axial symmetry, suggesting the existence of an extremum in between. 
Moreover, the maximum displacement reached in $u_r$ is found to be about 3 times larger in comparison to that reached in $u_\theta$.

By examining the displacement at various forcing frequencies, we observe that larger frequencies induce remarkably smaller deformation since the capsule membrane does not have enough time to respond to the fast oscillating particle.
{In typical situations, the forces acting by optical tweezers on suspended particles are of the order of 1~pN \cite{cipparrone10} and the capsule has a radius $10^{-6}~$m and a shearing modulus $5 \times 10^{-6}~$N/m \cite{Freund_2014}. 
For a forcing frequency $\beta = 4$, the membrane undergoes a maximal deformation of about 1~\% of its undeformed radius.
Therefore, deformations are significantly small and deviations from sphericity are negligible. 
The analytical predictions based on the linear theory of small deformation are found to be in a good agreement with simulations.
A small deviation is observed notably for $u_\theta$ at small angles which is possibly due to a finite size effect since the analytical predictions are based on the point-particle approximation whereas simulations treat truly extended particle of finite size. }

\section{Conclusion}\label{sec:conclusions}

Using the image solution technique, we have computed the leading-order hydrodynamic self-mobility of a solid spherical particle axisymmetrically moving nearby a large deformable capsule whose membrane exhibits resistance towards shearing and bending.
The mobility corrections are expressed in terms of infinite but convergent series whose coefficients are frequency-dependent complex quantities.
We have shown that in the vanishing frequency limit, the particle self-mobility near a hard sphere is recovered only when the membrane possesses a resistance towards shearing.
For a large membrane radius, our results perfectly overlap with those obtained earlier for a planar membrane in the high frequency regime.
The major qualitative difference between the planar and the spherical membrane is the existence of a second, low-frequency peak in the imaginary part (and a corresponding dispersion step in the real part) caused by shear resistance.
{
The appearance of two peaks can be understood by the simple fact that the membrane traction jump stemming from shear resistance contains contributions from normal (radial) as well as in-plane (tangential) displacements.
For a planar membrane, only in-plane displacements contribute to shear resistance which explains why the observed peak disappears at large radii.}
For a bending-only membrane, curvature effects are much less pronounced and the planar membrane gives a fairly good approximation even deep in the low frequency regime.

Considering the capsule motion, we have found that the pair-mobility function depends solely on the membrane shearing properties and it can be well described by a Debye-like model with a single relaxation time.
The pair-mobility function for a bending-only membrane is therefore frequency-independent and it is identical to that for a viscous drop.
We have further found that the point particle approximation despite its simplicity leads to a very good agreement with the numerical simulations preformed for a truly extended particle using a completed double layer boundary integral method.

\acknowledgments
The authors thank the Volkswagen Foundation for financial support and acknowledge the Gauss Center for Supercomputing e.V. for providing computing time on the GCS Supercomputer SuperMUC at Leibniz Supercomputing Center. 
We thank Achim Guckenberger and Maciej Lisicki for useful discussions and technical support.

\appendix

\section{Membrane mechanics}\label{appendix:membraneMechanics}

In this appendix, we shall derive equations in spherical coordinates for the traction jump across a membrane endowed with shearing and bending rigidities.
Here we follow the convention  in which the symbols for the radial, azimuthal and polar angle coordinates are taken as $r$, $\phi$ and $\theta$ respectively,
with the corresponding orthonormal basis vectors $\eR$, $\ePhi$ and $\eThe$.
Similar, all the lengths will be scaled by capsule radius $a$.
We denote by $\vect{a} = \eR$ the position vector of the points located at the undisplaced membrane.
After axisymmetric deformation, the vector position reads
\begin{equation}
 \vect{r} = (1+u_r)\eR + u_\theta \eThe  \, ,
\end{equation}
where $u_r$ and $u_\theta$ denote the radial and meridional displacements.
In the following, capital Roman letters shall be reserved for the undeformed state while small letters for the deformed.
The spherical membrane can be defined by the covariant base vectors $\gOne := \vect{r}_{,\theta}$ and $\gTwo := \vect{r}_{,\phi} $.
The unit normal vector $\vect{n}$ is defined in such a way to form a direct trihedron with $\gOne$ and $\gTwo$.
The covariant base vectors are
\begin{subequations}
 \begin{align}
  \gOne &=  (u_{r,\theta} - u_{\theta}) \eR + (1 + u_r + u_{\theta,\theta} ) \eThe \, , \\
  \gTwo &=  \big( (1 + u_r) \sin \theta + u_{\theta} \cos \theta \big) \ePhi \, ,
 \end{align}
\end{subequations}
and the unit normal vector at leading order in deformation reads
\begin{equation}
 \vect{n} = \eR - \left( u_{r,\theta} - u_{\theta} \right) \eThe  \, . 
\end{equation}

Note that $\gOne$ and $\gTwo$ have (scaled) length dimension while $\vect{n}$ is dimensionless.
In the deformed state, the components of the metric tensor are defined by the scalar product $g_{\alpha\beta} = \vect{g}_{\alpha} \cdot \vect{g}_{\beta}$.
The contravariant tensor $g^{\alpha\beta}$, defined as the inverse of the metric tensor, is linearized as
\begin{equation}
 g^{\alpha\beta} = \left(
                   \begin{array}{cc}
                    1 - 2\epsilon_{\theta\theta} & 0 \\
                    0 & \frac{1-2\epsilon_{\phi\phi}}{\sin^2 \theta}
                   \end{array}
		   \right) \, ,
		    \label{contravariantTensor}
\end{equation}
where $\epsilon_{\alpha\beta}$ represents the components of the in-plane strain tensor written in spherical coordinates as \cite{sadd09}
\begin{subequations}
 \begin{align}
  \epsilon_{\theta\theta} &= u_r + u_{\theta,\theta} \, , \\
  \epsilon_{\phi\phi} &=  u_r + u_{\theta} \cot \theta \, .
 \end{align}
\end{subequations}

The contravariant tensor in the undeformed state $G^{\alpha\beta}$ can immediately be obtained by considering a vanishing strain tensor in Eq.~\eqref{contravariantTensor}.

\subsection{Shearing contribution}

In this subsection, we shall derive the traction jump equations across a membrane endowed with an in-plane shearing resistance.
The two invariants of the strain tensor are given by Green and Adkins as \cite{green60, zhu14}
\begin{subequations}
 \begin{align}
  I_1 &= G^{\alpha\beta} g_{\alpha\beta} - 2 \, , \\
  I_2 &= \det G^{\alpha\beta} \det g_{\alpha\beta} - 1 \, .
 \end{align}
\end{subequations}

The contravariant components of the stress tensor $\tau^{\alpha\beta}$ can then be obtained provided knowledge of the membrane constitutive elastic law, whose areal strain energy functional is $W (I_1,I_2)$, such that \cite{lac04}
\begin{equation}
 \tau^{\alpha\beta} = \frac{2}{\JS} \frac{\partial W}{\partial I_1} G^{\alpha\beta} + 2\JS \frac{\partial W}{\partial I_2} g^{\alpha\beta} \, ,
 \label{stressTensor}
\end{equation}
where $\JS := \sqrt{1+I_2}$ is the Jacobian determinant, prescribing the ratio between deformed and undeformed local areas.
In the linear theory of elasticity, $\JS \simeq 1 + e$, where $e := \epsilon_{\theta\theta} + \epsilon_{\phi\phi}$ being the trace of the in-plane strain tensor, commonly know as the dilatation.
In this work, we use the Skalak model to describe the elastic properties of the capsule membrane, whose areal strain energy reads \cite{krueger12, Freund_2014}
\begin{equation}
 W(I_1, I_2) = \frac{\kS}{12} \left( I_1^2 + 2I_1-2I_2 + C I_2^2 \right) \, ,
 \label{skalakEquation}
\end{equation}
where $C:=\kA/\kS$.
Note that for $C=1$, the Skalak model is equivalent to the Neo-Hookean model for small deformations \cite{lac04}.
After plugging Eq.~\eqref{skalakEquation} into Eq.~\eqref{stressTensor}, the linearized in-plane stress tensor reads
\begin{equation}
 \tau^{\alpha\beta} = \frac{2 \kS}{3}
 \left(
 \begin{array}{cc}
  \epsilon_{\theta\theta} + Ce & 0 \\
  0 & \frac{\epsilon_{\phi\phi} + Ce}{\sin^2 \theta} 
 \end{array}
 \right) \, .
\end{equation}

The membrane equilibrium equations balancing the elastic and  external forces read
\begin{subequations}
 \begin{align}
  \nabla_{\alpha} \tau^{\alpha\beta} + \Delta f^{\beta} &= 0 \, , \label{Equilibrium_Tangential} \\
  \tau^{\alpha\beta} b_{\alpha\beta} + \Delta f^{n} &= 0 \, , \label{Equilibrium_Normal}
 \end{align}
\end{subequations}
where $\Delta \vect{f} = \Delta f^{\beta} \vect{g}_{\beta} + \Delta f^{n} \vect{n} $ is the traction jump across the membrane and $\nabla_{\alpha}$ denotes the covariant derivative defined for a second-rank tensor as
\begin{equation}
 \nabla_{\alpha} \tau^{\alpha\beta} = \tau^{\alpha\beta}_{,\alpha} + \Gamma_{\alpha\eta}^{\alpha} \tau^{\eta\beta} + \Gamma_{\alpha\eta}^{\beta} \tau^{\alpha\eta} \, ,
\end{equation}
and $\Gamma_{\alpha\beta}^{\lambda}$ are the Christoffel symbols of the second kind defined as \cite{synge69} [ch. 2]
\begin{equation}
 \Gamma_{\alpha\beta}^{\lambda} = \frac{1}{2} g^{\lambda\eta} \left( g_{\alpha\eta,\beta} + g_{\eta\beta,\alpha} - g_{\alpha\beta, \eta} \right) \, .
\end{equation}

Continuing, $b_{\alpha\beta}$ is the second fundamental form (curvature tensor) defined as
\begin{equation}
 b_{\alpha\beta} = \vect{g}_{\alpha,\beta} \cdot \vect{n} \, .
\end{equation}

Note that at zeroth order, the non-vanishing components of the Christoffel symbols are $\Gamma_{\phi\theta}^{\phi} = \Gamma_{\theta\phi}^{\phi} = \cot \theta$ and $\Gamma_{\phi\phi}^{\theta} = -\sin\theta\cos\theta$.
After some algebra, we find that the meridional tangential traction jump across the membrane given by Eq.~\eqref{Equilibrium_Tangential} reads
\begin{equation}
   \tau^{\theta\theta}_{,\theta}  + \Gamma_{\phi\theta}^{\phi} \tau^{\theta\theta} + \Gamma_{\phi\phi}^{\theta} \tau^{\phi\phi} + \Delta f^{\theta} = 0 \, .
\end{equation}

At zeroth order, the non-vanishing components of the curvature tensor are $b_{\theta\theta} = -1$ and $b_{\phi\phi} = -\sin^2\theta$.
For the normal traction jump Eq.~\eqref{Equilibrium_Normal} we therefore get
\begin{equation}
 -\tau^{\theta\theta} - \sin^2 \theta \tau^{\phi\phi} + \Delta f^{n} = 0 \, .
\end{equation}

After substitution and writing the projected equations in the spherical coordinates basis vectors, we immediately get the following set of equations,
\begin{subequations}
 \begin{align}
   \frac{2\kS}{3} & \bigg( (1+C)\epsilon_{\theta\theta,\theta} + C\epsilon_{\phi\phi,\theta} \notag \\
		&+ (\epsilon_{\theta\theta} - \epsilon_{\phi\phi}) \cot \theta \bigg) 
		+ \Delta f_{\theta} = 0 \, ,  \\
                -\frac{2\kS}{3} & (1+2C) \left( \epsilon_{\theta\theta} + \epsilon_{\phi\phi} \right) + \Delta f_n = 0 \, .
 \end{align}
 \label{equilibriumEqn}
\end{subequations}

We further mention that for curved membranes, the normal traction jump does not vanish in the \emph{plane stress} formulation employed here because the zeroth order in the curvature tensor is not identically null.
Indeed, this is not the case for a planar elastic membrane where the resistance to shearing only introduces a jump in the tangential traction jumps  \cite{daddi16, daddi16b}.

By substituting $\epsilon_{\theta\theta}$ and $\epsilon_{\phi\phi}$ with their expressions,  Eqs.~\eqref{equilibriumEqn} turn into the traction jumps equations \eqref{tractionJumpShear}.


\subsection{Bending contribution}

For the bending resistance, we use the linear model, in which the bending moment is related to the curvature tensor via \cite{pozrikidis01, zhu15}
\begin{equation}\label{eq:bending}
 M_{\alpha}^{\beta} = -\kB \left( b_{\alpha}^{\beta} - B_{\alpha}^{\beta} \right) \, ,
\end{equation}
where $\kB$ is the bending modulus and the spontaneous curvature is set to $B_\alpha^\beta=\vect{G}_{\alpha,\beta} \cdot \vect{n}$ corresponding to the undeformed sphere.
The mixed version of the curvature tensor~$b_{\alpha}^{\beta}$ is related to the covariant representation via $b_{\alpha}^{\beta} = b_{\alpha\delta} g^{\delta\beta}$.
The contravariant components of the transverse shearing vector $\vect{Q}$ is obtained from a local torque balance with the applied moment as $Q^{\beta} = \nabla_{\alpha} M^{\alpha\beta}$.
Note that the raising and lowering indices operations imply that $M^{\alpha\beta} = g^{\alpha\gamma} g^{\beta\delta} M_{\gamma\delta}$ and that $M_{\alpha\beta} = M_{\alpha}^{\delta} g_{\delta \beta}$.
The meridional force reads
\begin{equation}
  Q^{\theta} = -{\kB} \Big( \left(1-\cot^2\theta\right)u_{r,\theta} + u_{r,\theta\theta} \cot\theta + u_{r,\theta\theta\theta} \Big)  \, . \notag
\end{equation}

The membrane equilibrium equations balancing the bending forces reads
\begin{subequations}
 \begin{align}
  -b_{\alpha}^{\beta} Q^{\alpha} + \Delta f^{\beta} &= 0 \, , \\
  \nabla_{\alpha} Q^{\alpha} + \Delta f^{n} &= 0 \, ,
 \end{align}
\end{subequations}
where for a first-rank tensor (vector) the covariant derivative is defined as $\nabla_{\beta} Q^{\alpha} = \partial_{\beta} Q^{\alpha} + \Gamma_{\beta\delta}^{\alpha} Q^{\delta}$.
The equilibrium equations can thus be written as
\begin{subequations}
 \begin{align}
  &-{\kB} \left( \left(1-\cot^2\theta\right)u_{r,\theta} + u_{r,\theta\theta} \cot\theta + u_{r,\theta\theta\theta} \right) + \Delta f_{\theta} = 0 \, , \label{Delta_f_phi_B} \\
  &-{\kB} \bigg( \left( 3\cot\theta+\cot^3\theta \right) u_{r,\theta} -  u_{r,\theta\theta}\cot^2\theta \notag \\
  &+ 2 u_{r,\theta\theta\theta}\cot\theta + u_{r,\theta\theta\theta\theta} \bigg) + \Delta f_{n} = 0 \,  \label{Delta_f_r_B}
 \end{align}
 \label{Delta_f_bending} 
\end{subequations}
corresponding to the traction jump given in Eq.~\eqref{tractionJumpBend}.


\section{Transformation equations between the scaled and physical quantities}\label{appendix:transformationEquations}

In this appendix, we shall state the transformation relations between the scaled and physical quantities.
The physical quantities are denoted by a tilde while the absence of tilde refers to the scaled ones.
For the variables with the dimension of length, such as $r$ and $R$, we have $\tilde{r} = r a$ and $\tilde{R}= R a$.
For the velocity we have $\tilde{v} = v a$, for the force $\tilde{F}=F a$, for the fluid viscosity $\tilde{\eta} = \eta /a$, for the pressure $\tilde{p}= p/a$ and similar for the traction jump $\tilde{\Delta f} = \Delta f / a$.
For the shearing modulus $\tilde{\kS} = \kS$, for the bending modulus $\tilde{\kB} = \kB a^2$.
It follows that $\tilde{\alpha}=\alpha a$ and $\tilde{\alphaB} = \alphaB a^3$.

\section{Force-free condition}\label{appendix:forceFree}

In this appendix, we shall show that for finite shearing modulus, the force free condition assumed for the capsule is satisfied.

The induced hydrodynamic force on the capsule is computed by integrating the normal stress vector over the capsule's outside surface $A^{+}$ as \cite{felderhof14}
\begin{equation}
 \F_1 = \int_{A^{+}} \boldSigma \cdot \eR \, \Intd A = A_0 \Fext  \, ,
\end{equation}
meaning that the hydrodynamic force in the multipole expansion is given only by the coefficient of the monopole field \cite{kim13}.
For shearing-only and bending-only membranes, we have shown that $A_0 = 0$ as can be inferred from Eqs.~\eqref{A_n_Shearing} and \eqref{A_n_Bending}.
This is the case also for a membrane with both shearing and bending resistances.
We therefore conclude that no net force is exerted on the capsule.

We note that, for infinite shearing modulus, i.e. in the hard-sphere limit, $A_0 \ne 0$ as can clearly be seen in Eq.~\eqref{A_n_hardSphere}.
Additional singularities therefore need to be added to the reflected flow field in order to ensure the force free assumption (see Ref. \onlinecite{fuentes88} for further details.)

\section{Estimation of the number of terms required for the computation of particle self-mobility}\label{appendix:seriesEstimation}

In this appendix, we shall determine the number of terms required for the computation of particle self-mobility in order to achieve a given precision.

Let us denote by $f_n (\xi)$ the general term of the function series giving the particle mobility correction in Eq.~\eqref{mobilityCorrection}.
For a large value of $n$, we have the leading order asymptotic behavior
\begin{equation}
 f_n(\xi) = \frac{3b}{8}  \left( 1-\xi^2 \right)^2 n^2 \xi^{2n+4} + \bigO \left( n \xi^{2n} \right) \, , \label{f_n} %
\end{equation}
which does not depend on capsule shearing and bending properties. 
In order to compute an infinite series numerically up to a given precision, we define the truncation error as
\begin{equation}
 \begin{split}
  E(N) &:= \left| \sum_{n=N+1}^{\infty} f_n(\xi) \right| \\
       &\simeq \frac{3b}{8} \frac{-N^2 \xi^4+(2N^2+2N-1)\xi^2-(N+1)^2}{1-\xi^2} \xi^{2N+6} \, . \notag
 \end{split}
\end{equation}

Given a certain precision $\varepsilon$, the number of terms $N$ required to achieve the desired precision can be determined by solving the inequality 
$E(N) < \varepsilon$.
For example, by taking $h=2b$, $b=1/10$ and requiring a precision $\varepsilon = 10^{-4}$, only 29 terms in the series are needed.
For $b=10^{-3}$ however, 2993 terms are needed.
As a result, more terms are required for convergence when the capsule radius is taken very large, i.e. when $\xi\sim 1$.
By requiring a precision $\varepsilon = 10^{-6}$, 44 and 4316 terms are necessary for $b=1/10$ and $b=10^{-3}$ respectively. 
A precision of $\varepsilon = 10^{-4}$ has been consistently adopted throughout this work.

\input{main.bbl}

\end{document}

%% file: main.bbl
%